\documentclass[twocolumn,twocolappendix]{aastex63}

\newcommand{\Att}{{\cal A}}

\received{}
\revised{}
\accepted{}
\submitjournal{ApJ}

\shorttitle{Estimating dust attenuation}
\shortauthors{Li et al.}

\begin{document}

\title{Estimating dust attenuation from galactic spectra. I. methodology and tests}

\correspondingauthor{Cheng Li}
\email{cli2015@tsinghua.edu.cn, liniu14@mails.tsinghua.edu.cn}

\author[0000-0002-0656-075X]{Niu Li}
\affiliation{Department of Astronomy, Tsinghua University, Beijing 100084, China}

\author[0000-0002-8711-8970]{Cheng Li}
\affiliation{Department of Astronomy, Tsinghua University, Beijing 100084, China}

\author{Houjun Mo}
\affiliation{Department of Astronomy, Tsinghua University, Beijing 100084, China}
\affiliation{Department of Astronomy, University of Massachusetts Amherst, MA 01003, USA}

\author{Jian Hu}
\affiliation{National Astronomical Observatories, Chinese Academy of Sciences, Datun Road 20A, Beijing 100012, China}
\affiliation{Shenzhen Middle School, Shenzhong Street 18, Shenzhen 518001, China}

\author{Shuang Zhou}
\affiliation{Department of Astronomy, Tsinghua University, Beijing 100084, China}

\author{Cheng Du}
\affiliation{Department of Astronomy, Tsinghua University, Beijing 100084, China}


\begin{abstract}
  We develop a method to estimate the dust attenuation curve of
  galaxies from full spectral fitting of their optical
  spectra. Motivated from previous studies, we separate the
  small-scale features from the large-scale spectral shape, by
  performing a moving average method to both the observed spectrum
  and the simple stellar population model spectra. The intrinsic
  dust-free model spectrum is then derived by fitting the observed
  ratio of the small-scale to large-scale (S/L) components with the
  S/L ratios of the SSP models. The selective dust attenuation curve
  is then determined by comparing the observed spectrum with the
  dust-free model spectrum.  One important advantage of this method
  is that the estimated dust attenuation curve is independent of the
  shape of theoretical dust attenuation curves. We have done a
  series of tests on a set of mock spectra covering wide ranges of
  stellar age and metallicity. We show that our method is able to
  recover the input dust attenuation curve accurately, although the
  accuracy depends slightly on signal-to-noise ratio of the spectra.
  We have applied our method to a number of edge-on galaxies with
  obvious dust lanes from the ongoing MaNGA survey, deriving their
  dust attenuation curves and $E(B-V)$ maps, as well as dust-free
  images in $g$, $r$, and $i$ bands. These galaxies show obvious
  dust lane features in their original images, which largely
  disappear after we have corrected the effect of dust
  attenuation. The vertical brightness profiles of these galaxies
  become axis-symmetric and can well be fitted by a simple model
  proposed for the disk vertical structure.
  Comparing the estimated dust attenuation curve
  with the three commonly-adopted model curves, we find that 
  the Calzetti curve provides the best description of the 
  estimated curves for the inner region of galaxies, while the 
  Milky Way and SMC curves work better for the outer region.

\end{abstract}



\section{Introduction} \label{sec:intro}

The observed spectrum of a galaxy is a combination of several components: 
a continuum, absorption and emission lines. The continuum and absorption 
lines are both dominated by starlight, thus usually referred to as the 
stellar component of the spectrum, 
while the emission-line component is produced in H{\sc ii} regions around 
hot stars, or emission-line regions of active nuclei, or both. 
All these components, 
however, are modified by the attenuation of dust grains 
distributed in the inter-stellar space. Dust attenuation can affect galaxy spectra 
over a wide range of wavelengths, from ultraviolet (UV), optical to infrared, 
by absorbing short-wavelength photons in UV/optical and re-emitting 
photons in the infrared, and the absorption is stronger in shorter wavelength.
Consequently, dust attenuation can cause changes in the overall shape of a galaxy 
spectrum. Such attenuation has to be taken into account before one can measure 
the different components of an observed spectrum reliably.

Various schemes have been used to estimate dust
  attenuation.  For Milky Way and nearby galaxies,
  such as the Magellanic Clouds, observations of individual stars can
  directly probe the dust extinction along different lines of sight
  \citep[e.g.,][]{1984A&A...132..389P,1989ApJ...345..245C,1999PASP..111...63F,2003ApJ...594..279G}.
  Far-infrared (FIR) observations provide direct measurements of dust
  attenuation for more distant galaxies because the emission by dust
  dominates the spectral energy distribution (SED) of FIR, although
  such observations can be made only  through space telescopes. When
  both infrared (IR) and UV photometry are available,  dust
  attenuation may be estimated by the IR-to-UV luminosity ratio, $\rm
  L_{IR} / L_{UV}$, known as IRX, as described in
  \citep[e.g.,][]{1999ApJ...521...64M,2000ApJ...533..236G}.  In the
  absence of IR data, the slope of the UV continuum spectrum $\beta$
  can also be used as an alternative estimator
  \citep[e.g.,][]{1994ApJ...429..582C,1999ApJ...521...64M}.  When
  multi-band photometry covering a wide range of wavelengths is
  available,  a commonly used method is to fit the observed SED with
  stellar population synthesis models, which provide estimates of  a
  variety of stellar population parameters including dust attenuation
  \citep[e.g.,][]{2009A&A...507.1793N,2013MNRAS.436.2535J,2016MNRAS.462.1415C}.

  The spectrum of a galaxy should, in principle, contain more detailed
  information about the physical properties of the galaxy.  For
  star-forming regions which are usually dusty, the attenuation  of
  various lines is commonly estimated from the Balmer decrement,  by
  comparing the observed $\rm H_\alpha$ to $\rm H_\beta$ line ratio
  ($\rm H_\alpha/H_\beta$) with the intrinsic one predicted by atomic
  physics applied to a given environment \citep{2006agna.book.....O}.
  Dust attenuation in the stellar component, however, cannot be
  estimated in the same way due to at least two factors.  First, the
  attenuation by star-forming regions and that  by stars can be quite
  different, with the former  expected to be stronger in most
  cases. Second, the Balmer decrement is not measurable in non-star
  forming regions where emission lines are weak.  Therefore, dust
  attenuation in the stellar component is usually estimated through
  full spectral fitting. For instance, a simple approach is to match a
  dust-reddened galactic spectrum with its un-reddened  counterpart
  that is produce by a similar stellar population
  \citep[e.g.,][]{2000ApJ...533..682C,2011MNRAS.417.1760W,2015ApJ...806..259R,
    2017ApJ...840..109B,2017ApJ...851...90B}.

  Alternatively,  the
  estimate of the stellar dust attenuation is made by fitting the full
  spectrum with a stellar population synthesis model
  \citep[e.g.,][]{2013ARA&A..51..393C,2015MNRAS.449..328W,2018A&A...613A..13Y,2018MNRAS.478.2633G}.
  A number of spectral synthesis fitting codes are now publicly
  available and are  commonly adopted for such purposes. These include
  \texttt{PPXF} developed by
  \citealt{2004PASP..116..138C,2017MNRAS.466..798C},
  \texttt{STARLIGHT} by \citealt{2005MNRAS.358..363C},
  \texttt{STECKMAP} by
  \citealt{2006MNRAS.365...46O,2006MNRAS.365...74O},  and
  \texttt{VESPA} by \citealt{2007MNRAS.381.1252T}.  In the
  conventional fitting method, one usually makes an assumption  about
  the shape of dust attenuation curve and treats the amount of
  attenuation as an free parameter to be obtained from the fitting.
  Obviously, the fitting result will depend on the theoretical dust
  attenuation curve chosen. In addition,  it is known that dust
  attenuation has significant degeneracy with stellar age in such
  fitting. 

In order to overcome some of the problems in the conventional method,  
\citet{2015MNRAS.449..328W,2017MNRAS.472.4297W} have developed 
a code of full spectral fitting (\texttt{Firefly}), using a new 
approach to break the degeneracy between dust and other stellar 
population properties. In this method, dust attenuation is assumed to
affect the large-scale shape of an observed spectrum but little 
the spectral shapes on small scales. Their solution is to first apply 
a high-pass filter (HPF) to remove the large-scale features in both 
the observed and model spectra, before fitting the filtered observed 
spectrum with the filtered model spectra.

In this paper we adopt a method similar to that of 
\citet{2015MNRAS.449..328W,2017MNRAS.472.4297W} to fit the 
spatially resolved spectra of edge-on disk galaxies selected 
from SDSS IV MaNGA \citep{2015ApJ...798....7B, 2017AJ....154...28B} to investigate the dust attenuation
maps of these galaxies. To do this, we separate 
small-scale spectral features from large-scale variations 
using a moving box average rather than the Fourier transform
adopted in \citet{2015MNRAS.449..328W,2017MNRAS.472.4297W}. 
Since at a given wavelength dust attenuation affects both the  
small-scale and large-scale components in a similar way, 
the ratio between them is expected to be independent of dust 
attenuation, and so can be used to constrain the underlying 
stellar population. The paper is organized as follows. \S\ref{sec:method} 
describes our method. In \S\ref{sec:mock}, we test the performance 
of our method with mock spectra. We then apply our method to the MaNGA data in 
\S\ref{sec:manga} and compare our results with those obtained earlier. 
Finally, we summarize and discuss in \S\ref{sec:summary}.

\section{The method} \label{sec:method}
A simple method to extract information from an observed galaxy spectrum
is to fit it to a linear combination of simple stellar populations (SSPs). 
Each SSP is a single, coeval population of stars with a given metallicity 
and abundance pattern. The spectral synthesis of a SSP, therefore, 
consists of three components: the evolution of individual stars 
in the form of isochrones; a library of stellar spectra; and an initial 
mass function (IMF). Mathematically, the spectrum of a SSP of metallicity 
$Z$ at the age $t$ can be written as 
\begin{small}
\begin{equation} \label{eq:ssp}
  f_{\rm SSP}(t,Z) = \int_{m_0}^{m(t)}
  f_{\rm star}\left[T_{\rm eff}(M),\log g(M)\vert t, Z \right]
  \Phi (M)\,{\rm d}M,
\end{equation}
\end{small}
where $f_{\rm star}$ is the spectrum of a star of age $t$, metallicity $Z$
and initial mass $M$ in the spectral library, and $\Phi(M)$ is the IMF
\citep[e.g.,][]{2013ARA&A..51..393C}.
The dependence of the effective temperature, $T_{\rm eff}$, 
and the surface gravity, $g$, on the initial stellar mass for given 
$t$ and $Z$ is determined by the stellar evolution model adopted. 
The lower limit of integration, $m_0$, is typically taken 
to be the hydrogen burning limit, $0.08 M_{\sun}$, and the upper 
limit, $m(t)$, is the mass of the most massive stars that can survive
to the age $t$, as determined by the stellar evolution model.

There are several popular stellar population synthesis codes available
\citep[e.g.,][]{1999ApJS..123....3L,2003MNRAS.344.1000B,2005MNRAS.362..799M,2010MNRAS.404.1639V}. 
In this paper, we use the simple stellar populations 
given by \citet[][hereafter BC03]{2003MNRAS.344.1000B}.
Using the Padova 1994 evolutionary tracks and Chabrier initial mass function \citep{2003PASP..115..763C},
BC03 provides a large sample of SSPs, covering 221 ages from $t=0$ years to 
$t=2.0\times 10^{10}$ years, and six metallicities from $Z=0.0001$ 
to $Z=0.05$ (note that the solar metallicity $Z_{\sun}=0.02$) at a 
spectral resolution of 3\AA. A total of 1326 SSPs are provided by BC03.

Once we have a series of SSPs with different ages and metallicities, 
the observed spectrum, $F_{\rm obs}(\lambda)$, can be fitted with a linear 
combination of the SSPs together with a model of dust attenuation:
\begin{eqnarray}\label{eq:sspfit}
  F^{\rm a}(\lambda) &=& F(\lambda ) \cdot \Att(\lambda),\nonumber\\ 
  F(\lambda ) &\equiv& \sum\limits_{j = 1}^{N_ *} x_j f_{\rm SSP}^j(\lambda),
\end{eqnarray}
where $N_*$ is the number of templates (SSPs in our case), $f_{\rm SSP}^j$ is 
the spectrum of the $j^{th}$ SSP, $x_j$ is the weight of the $j^{th}$ SSP, 
and $\Att(\lambda)$ is the dust attenuation curve. So defined, 
${F^{\rm a}}(\lambda )$ is the model spectrum that takes into account dust 
attenuation, and $F(\lambda)$ is the dust free model spectrum. 

\begin{figure*}
  \centering
  \includegraphics[width=0.95\textwidth]{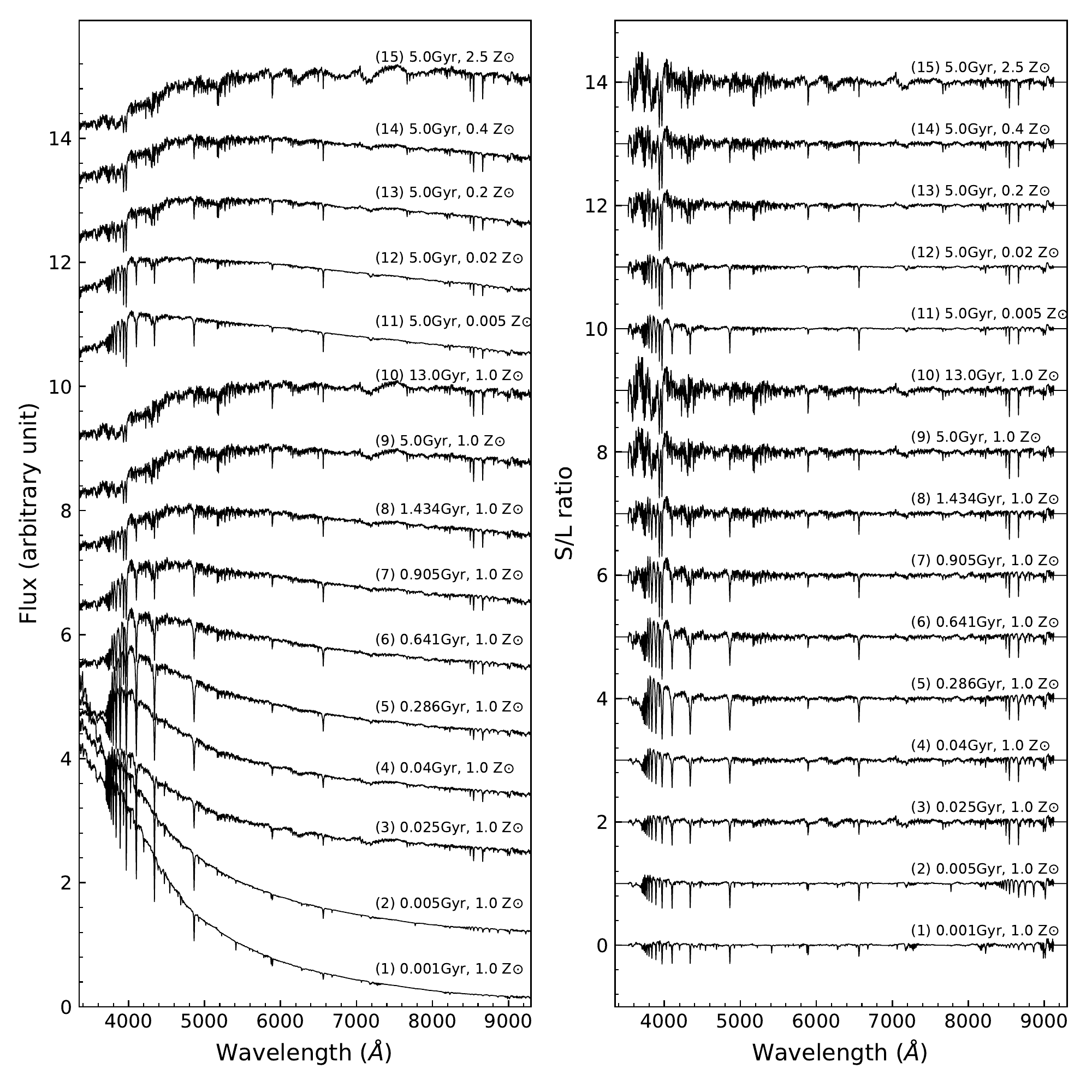}
  \caption{In the left panel, SSPs with different ages and metallicities are from BC03. The first 10 SSPs have different ages but the same metallicity (solar metallicity), and the last 5 SSPs have the same age (5 Gyr) but different metallicities. They are normalized at 5500\AA\ and the $i^{th}$ SSP has been shifted upward $i-1$ unites. The right panel shows the ratios of small-scale to large-scale component of each SSPs and the $i^{th}$ is shifted upward $i-1$ unites.}
  \label{fig:ssp_r}
\end{figure*}

\begin{figure*}
  \centering
  \includegraphics[width=0.95\textwidth]{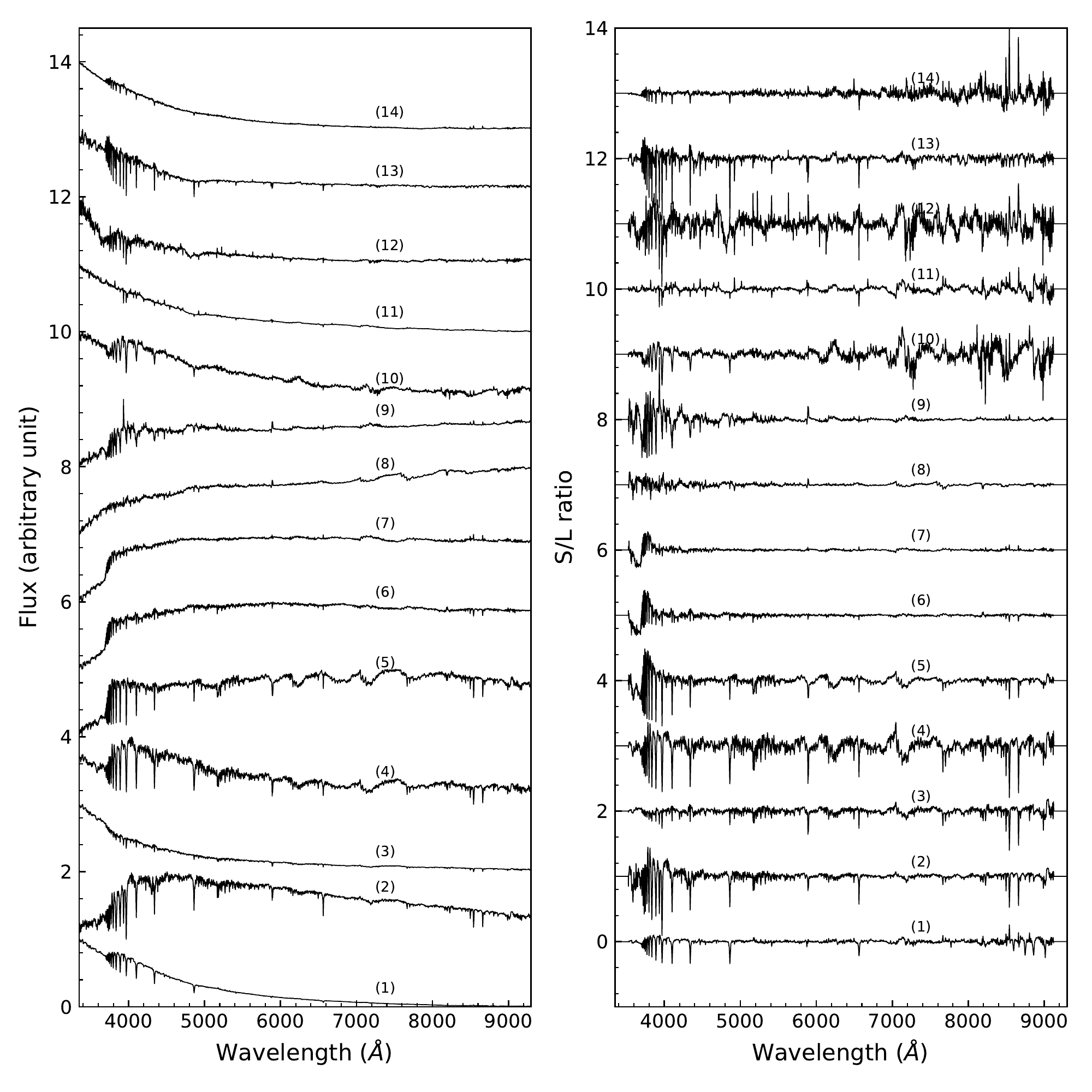}
  \caption{The left panel displays the 14 eign-spectra which are resulted from
  a Principal Component Analysis of the BC03 SSPs and used in this 
  work as the model templates. The right panel displays the corresponding
  ratio spectra between the small scale and large scale components.}
  \label{fig:ssp_pca}
\end{figure*}

\subsection{Spectral decomposition} \label{subsec:spec_dec}

In our analysis, we first decompose a spectrum into two components,
one small-scale component, $S$, and one large-scale component, $L$.
Roughly speaking, the $L$ component can be considered as the continuum
shape of  the spectrum, and the $S$ component as the composition of
absorption and emission line features.  We adopt a moving average
method to separate small-scale and large-scale
components. Specifically, the large-scale component is firstly
obtained through
\begin{equation}\label{eq:large}
  F_L^{\rm a}(\lambda )= \frac{1}{\Delta \lambda}
  \int_{\lambda-\Delta \lambda/2}^{\lambda+\Delta \lambda/2}
  F^{\rm a}(\lambda')\,d\lambda',
\end{equation}
where $\Delta \lambda$ is the size of the wavelength window. 
The small-scale component is simply defined as 
\begin{equation}\label{eq:small}
  F_S^{\rm a}(\lambda)=F^{\rm a}(\lambda)-F_L^{\rm a}(\lambda).
\end{equation}
Using equation (\ref{eq:sspfit}), we can write
\begin{eqnarray}\label{eq:large2}
  F_L^{\rm a}(\lambda )&=&\frac{1}{\Delta \lambda}
  \int_{\lambda-\Delta \lambda/2}^{\lambda+\Delta \lambda/2}
  F(\lambda') \cdot \Att(\lambda')\, d\lambda' \nonumber \\
  &\approx& \frac{\Att(\lambda)}{\Delta \lambda}
  \int_{\lambda-\Delta \lambda/2}^{\lambda+\Delta \lambda/2}
  F(\lambda')\,d\lambda',
\end{eqnarray}
where an approximation is made in the second line that
$\Delta \lambda$ is small and $\Att(\lambda)$ is smooth 
so that $\Att$ varies little over $\Delta \lambda$. It is then easy to 
see that
\begin{equation}\label{eq:large3}
  F_L^{\rm a}(\lambda )=F_L(\lambda) \cdot \Att(\lambda),
\end{equation}
where $F_L(\lambda)$ is the large-scale component of the intrinsic 
spectrum $F(\lambda)$. Similarly, equation (\ref{eq:small}) can be written as
\begin{eqnarray}\label{eq:small2}
  F_S^{\rm a}(\lambda )
  &=&\left[F(\lambda) - F_L(\lambda)\right] \cdot \Att(\lambda) \nonumber \\
  &=&F_S(\lambda) \cdot \Att(\lambda),
\end{eqnarray}
where $F_S(\lambda)$ is the small-scale component of the intrinsic spectrum. 
From equations (\ref{eq:large3}) and (\ref{eq:small2}), we see 
that the ratio between the small-scale and large-scale components is dust free:
\begin{equation}\label{eq:r_att}
  R^{\rm a}(\lambda)=\frac{F_S^{\rm a}(\lambda )}{F_L^{\rm a}(\lambda)}
  =\frac{F_S(\lambda)}{F_L(\lambda)}=R(\lambda),
\end{equation}
where $R(\lambda)$ is the ratio of the intrinsic spectrum.

As mentioned, our method of measuring the relative dust attenuation
curves was motivated from \citet{2015MNRAS.449..328W}, which
originally proposed the idea of decomposing an observed spectrum into
small- and large-scale components before performing full spectral
fitting. In that study, the decomposition was done by applying a
high-pass filter (HPF) to the Fourier
transform of the observed spectrum, thus removing large-scale
features through the use of an empirically chosen window function.
The filtered spectrum containing only small-scale features is then
fitted to the model templates that are filtered in the same way,
producing the best-fit, dust-free spectrum. 
This approach is not applicable to our method, however. First,
we need to obtain both the small-scale and large-scale components 
simultaneously, as we want to determine the best-fit stellar
spectrum by fitting the ratio between the small and large-scale components,
$R_{\lambda}$, instead of the filtered spectrum (i.e. the small-scale
component only) as in \citet{2015MNRAS.449..328W}.
Second, the effect of dust attenuation to the small-scale 
component is not zero, and it is the $S/L$ ratio that is 
not affected by the dust attenuation, as equation (\ref{eq:r_att}) shows.
Using equations (\ref{eq:large2}) and (\ref{eq:small2}), 
one can see that a moving average filter is ideal to meet these
conditions.

It must be realized that the above relations are derived under the 
assumption that all the stellar populations have the same attenuation
given by $\Att(\lambda)$. In reality, dust attenuation may be more 
complicated and, in particular, may be different for 
different stellar populations. Along this line, \citet{2000ApJ...539..718C} 
proposed a two-component dust model, in which old stars are assumed to be 
mixed uniformly with a diffuse dust distribution while young stars 
are assumed to be embedded in birth clouds of stars that 
have larger dust optical depth than the diffuse component. This model 
is motivated by the empirical relation between stellar and  
nebular dust attenuation \citep{2000ApJ...533..682C}.

More generally, suppose that the $j^{th}$ stellar population (not necessarily 
a SSP) has an attenuation curve of $\Att^j(\lambda)$, defined by 
\begin{equation}\label{eq:optdepth}
    \Att^j(\lambda) = e^{-\tau^j(\lambda)},
\end{equation}
where $\tau^j(\lambda)$ is the dust optical depth at wavelength $\lambda$, which
is different for different stellar populations.
The ratio between the small-scale 
and large-scale components can then be written as 
\begin{equation}\label{eq:multi1}
    R^{\rm a}(\lambda)=\frac{\sum_j x_j F^j_S(\lambda) e^{-\tau^j(\lambda)}}
    {\sum_j x_j F^j_L(\lambda)e^{-\tau^j(\lambda)}}\,,
\end{equation}
where $F^j_S$ and $F^j_L$ are the small-scale and large-scale 
spectrum of the $j^{th}$ population. 
This equation reduces to equation (\ref{eq:r_att}), 
as expected, if $\tau^j(\lambda)$ is the same for all `$j$'
(which is the case if the dust distribution is a screen in front 
of the galaxy, as assumed above), or the intrinsic 
spectra of different stellar populations are the same.

Consider another simple case in which $\tau^j(\lambda) \ll 1$. 
In this case, one can write
\begin{eqnarray}\label{eq:multi3}
    R^{\rm a}(\lambda)& \approx \frac{\sum_j x_j F^j_S(\lambda)
    [1-\tau^j(\lambda)]} {\sum_j x_j F^j_L(\lambda)[1-\tau^j(\lambda)]} \nonumber \\
    & = \frac{T_S(\lambda) \sum_j x_j 
    F_S^j(\lambda)}{T_L(\lambda) \sum_j x_j F^j_L(\lambda)},
\end{eqnarray}
where
\begin{equation}\label{eq:multi4a}
    T_S(\lambda) = 1 - \left<\tau\right>_S(\lambda) = 1 - \frac{\sum_j x_j F^j_S(\lambda) 
    \tau^j(\lambda)}{\sum_j x_j F^j_S(\lambda)}\,,
\end{equation}
and
\begin{equation}\label{eq:multi4b}
    T_L(\lambda) = 1 - \left<\tau\right>_L(\lambda) = 1 - \frac{\sum_j x_j F^j_L(\lambda) 
    \tau^j(\lambda)}{\sum_j x_j F^j_L(\lambda)}.
\end{equation}
Note that both $\left<\tau\right>_S(\lambda)$ and $\left<\tau\right>_L(\lambda)$ 
are weighted averages of $\tau^j(\lambda)$. 
We have estimated $T_S(\lambda) / T_L(\lambda)$ using realistic spectra
and attenuation curves, and found the ratio to depend only weakly on $\lambda$ 
for $\tau^j < 1$, so that $R^{\rm a}(\lambda)$ is again independent of dust attenuation. 
However, for $\tau^j > 1$, the ratio $R^{\rm a}(\lambda)$ can be affected significantly 
by the differences in dust attenuation among different stellar populations.
We thus conclude that, $R^{\rm a}(\lambda)$ can well reflect the properties of the 
underlying intrinsic spectrum under certain assumptions. Throughout this paper we  
assume $R^{\rm a}(\lambda) \approx R (\lambda)$.
Actually in previous studies, dust is commonly treated as a
screen in front of the whole galaxy or a given region of the galaxy,
so that  all the stellar populations in the galaxy/region have the
same attenuation. In reality this cannot be always true. 
We will come back and consider more general cases in the future.

\subsection{The fitting procedure} \label{subsec:spec_fit}

In our modeling of galaxy spectra, we fit the ratio 
$R^{\rm a}(\lambda)$ obtained from an observed spectrum with 
the corresponding ratio of the model spectrum. To this end, 
we decompose each model template to obtain its small-scale and 
large-scale components as defined by equations (\ref{eq:large}) 
and (\ref{eq:small}). For a given set of fitting coefficients, 
$\{x_j\}$, we obtain the model prediction of the ratio,  
\begin{equation}\label{eq:fit1}
  R^{\rm m}(\lambda)=\frac{f_0\sum_{j = 1}^{N_ * }x_jf_S^j(\lambda)}
  {f_0\sum_{j = 1}^{N_ * } x_jf_L^j(\lambda)},
\end{equation}
where $N_*$ is the number of templates, 
$f_S^j(\lambda)$ and $f_L^j(\lambda)$ are the
small-scale and large-scale components of the $j^{th}$
template, $f^j(\lambda)$, respectively. In the equation, 
the constant $f_0$ is the normalization of the model template, 
and is the same for all the $f^{j}_S(\lambda)$ and $f^{j}_L(\lambda)$.
The predicted ratio $R^{\rm m}(\lambda)$ is then compared with the observed 
ratio $R^{\rm a}(\lambda)$ to determine $\{x_j\}$. The fitting is carried out by using 
\texttt{MPFIT}, which is a non-linear Least-squares Fitting code 
in IDL \citep{2009ASPC..411..251M}. 
We should point out that, the fact that the normalization factor 
$f_0$ is cancelled out in equation~(\ref{eq:fit1}) indicates that 
this factor cannot be directly determined from this fitting procedure.
We will come back to this point later.

In principle, one can use all the SSPs from the BC03 library,
or a random subset, as the model templates for the above fitting 
process. SSPs with different ages and metallicities have 
different small-scale features. Similarly, the small-scale to large-scale 
(S/L) component ratios of different SSPs also have different features.  
This is shown in Figure~\ref{fig:ssp_r}, where the first 10 SSPs 
have different ages but the same metallicity (solar metallicity), 
while the last 5 have the same age (5 Gyr) but different metallicities. 
The left-hand panel shows the original SSPs and the right-hand
panel shows the corresponding S/L ratio spectra. 
It is obvious that different SSPs have different absorption 
line strengths and other broader features. It is these 
differences that allow us to derive constraints on the stellar 
populations from a galaxy spectrum.

In order to speed up the fitting process, 
we construct a small set of model templates by applying 
the technique of Principal Component Analysis (PCA, \citealt{1964MNRAS.127..493D}) 
to the BC03 SSP spectra. PCA can effectively reduce the size of the 
template library, as the principal features of the library can be
described by the first few eigen-spectra. For instance, 
\citet{2005AJ....129..669L} obtained galactic eigen-spectra using PCA 
and showed that the first nine eigen-spectra already provided  
the base to model the stellar spectra of the galaxies in 
Sloan Digital Sky Survey Data Release 1 
(SDSS DR1, \citealt{2003AJ....126.2081A}). 
In our analysis, we apply the PCA to the 1326 BC03 SSPs, and
we adopt the first 14 eigen-spectra resulted from the 
PCA as the model templates for our spectral fitting. 
The cumulative
contribution to variance by the first 14 eigen-spectra is 
99.985\%, indicating that they contain nearly all the information
of the original 1326 SSPs. 
Figure~\ref{fig:ssp_pca} displays the 14 eigen-spectra (the left
panel), as well as the corresponding S/L ratio spectra (the right 
panel). We should point out that the 
number of eigen-spectra is determined so as to 
include as much as possible information of the original
SSPs, and simultaneously to reduce as much as possible the 
computing time. We have repeated our tests to be presented below,
by adopting a different number of eigen-spectra, and found that
the spectral fitting is little affected as long as the majority
of the original information of the whole SSP library 
($>99\%$) is contained in the adopted eigen-spectra. 

Once the coefficients, $x_j$, are obtained from the fitting, 
the best fitting spectrum, $F^{\rm fit}$, which is expected to  
not contain the effect of dust attenuation, can be reconstructed
by 
\begin{equation}\label{eq:fit2}
  F^{\rm fit}(\lambda) = f_0 \sum_{j = 1}^{{N_ * }} x_j f^j(\lambda).
\end{equation}
Note that the normalization factor $f_0$ is unknown, and it is 
set to be arbitrary for the moment. This means that the fitting procedure
described above gives only the shape of the dust-free spectrum.
The shape of the dust attenuation curve can then be obtained by 
comparing the best fitting spectrum with the observed spectrum:
\begin{equation}\label{eq:fit3}
  \Att(\lambda) = \frac{F^{\rm a}(\lambda)}{F^{\rm fit}(\lambda)}.
\end{equation}
Conventionally, $\Att(\lambda)$ is written as
\begin{equation}\label{eq:fit4}
  \Att(\lambda) = 10^{-0.4 \cdot A(\lambda)},
\end{equation}
where $A(\lambda)$ is the dust attenuation at the wavelength of 
$\lambda$. In practice, we normalize both $F^{\rm a}(\lambda)$ and 
$F^{\rm fit}(\lambda)$ at a given wavelength, 
\begin{equation}
  \lambda_V = 5500\mbox{\AA}, 
\end{equation}
and we have
\begin{equation}\label{eq:fit5}
  A(\lambda) - A_V =
  2.5 \log_{10} \frac{\tilde{F}^{\rm fit}(\lambda)}{\tilde{F}^{\rm a}(\lambda)},
\end{equation}
where $A_V$ is the dust attenuation at $\lambda_V$, and
\begin{eqnarray}\label{eq:normalized}
  \tilde{F}^{\rm fit}(\lambda) & = F^{\rm fit}(\lambda)/F^{\rm fit}(\lambda_V), \nonumber \\
  \tilde{F}^{\rm a}(\lambda)   & = F^{\rm a}(\lambda)/F^{\rm a}(\lambda_V)
\end{eqnarray}
are the dust-free and the observed spectrum normalized at 
$\lambda_V$, respectively.
Note that the factor $f_0$ is implicitly contained in both $F^{\rm fit}(\lambda)$ and
$F^{\rm fit}(\lambda_V)$, and so is cancelled out due to the normalization.
Therefore, what we obtain from this fitting procedure is the 
relative dust attenuation curve, i.e. $A(\lambda)-A_V$, 
which is the dust attenuation as a function of wavelength relative 
to the attenuation at $\lambda_V$.

Equation~(\ref{eq:fit5}) shows that our method provides
a direct measurement of the relative dust attenuation curve, 
with no need to assume a functional form for the curve. In practice, 
however, it may still be desirable to use a parametric 
form to represent the curve. As shown below, our measurements of $A(\lambda)-A(V)$ 
can well be represented by a second-order polynomial 
in most (if not all) cases, 
\begin{equation}\label{eq:fit6}
  A(\lambda)-A_V = a_1 \lambda^{-1} + a_2 \lambda^{-2}. 
\end{equation}
The parameters $a_1$ and $a_2$ are to be obtained by fitting 
the above function to the $A(\lambda)-A_V$ measured by using
equation~(\ref{eq:fit5}). 

Given the measured $A(\lambda)-A_V$ and setting $\lambda=\lambda_B=4400$\AA, 
we therefore can get the $(B-V)$ color excess 
\begin{equation}\label{eq:ebv}
    E(B-V)=A_B - A_V,
\end{equation}
where $A_B$ is the dust attenuation at the wavelength, 
$\lambda_B$, corresponding to the $B$-band.
Similarly, the selective attenuation curve, defined as 
$[A(\lambda) - A_V]/[A_B - A_V]$, can also be obtained.  
In practice, dust attenuation is sometimes described by 
the total attenuation curve, defined as  
\begin{eqnarray}\label{eq:fit7}
  k(\lambda) &= \frac{A(\lambda)}{E(B-V)} = \frac{A(\lambda)} {A_B - A_V} \nonumber \\
  &=\frac{A(\lambda) - A_V} {A_B - A_V} + R_V,
\end{eqnarray}
where $R_V$ is the value of the total attenuation curve in 
the $V$-band, 
\begin{equation}\label{eq:rv}
  R_V = A_V / E(B-V).
\end{equation}
$R_V$ is also known as the ratio of the total to selective 
attenuation in the $V$-band. 

\begin{figure*}
  \centering
  \includegraphics[width=0.95\textwidth]{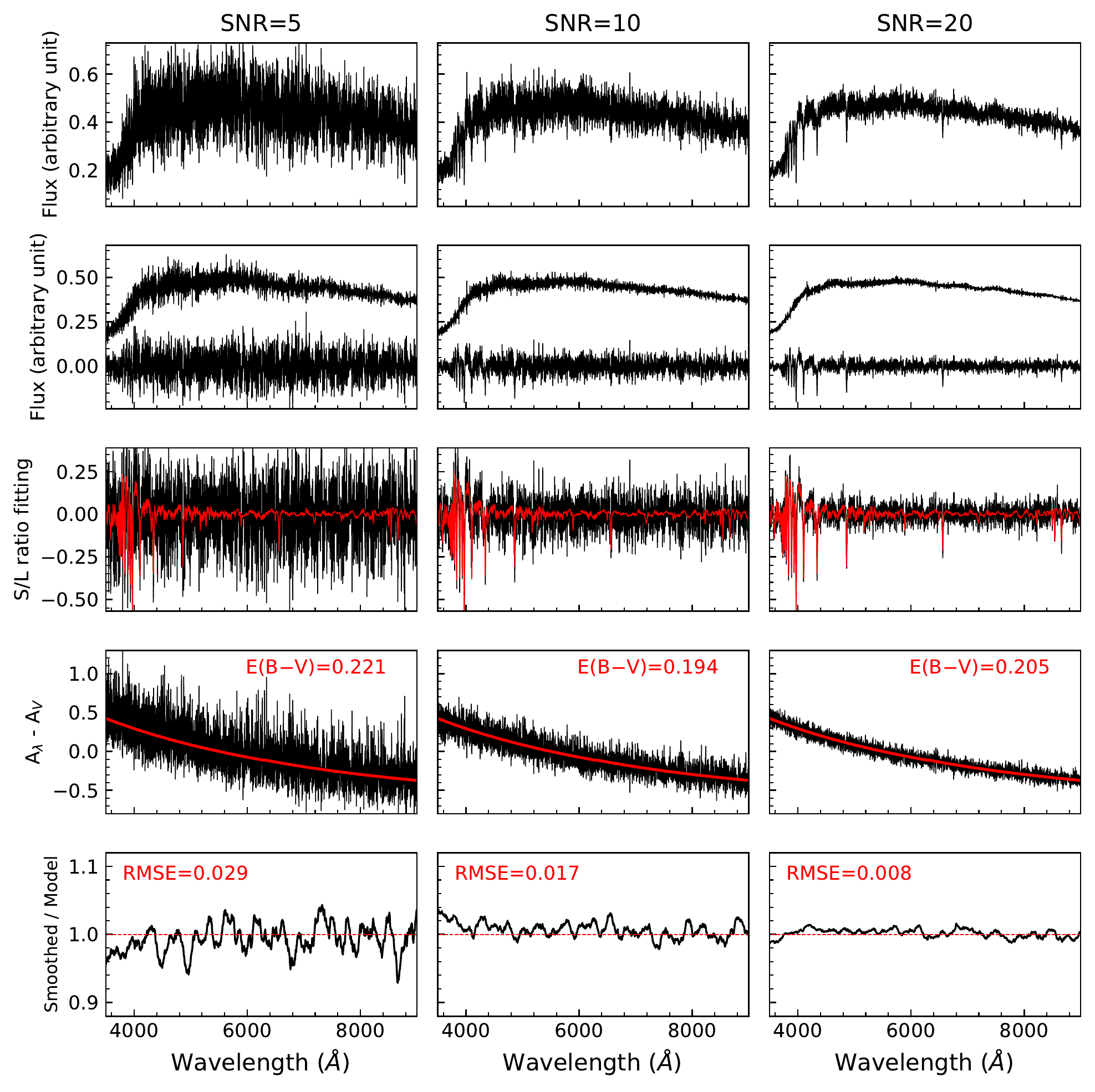}
  \caption{Examples of the fitting procedure for measuring relative dust 
  attenuation curves and $E(B-V)$. Panels from left to right correspond 
  to three mock spectra with different signal-to-noise ratios (SNRs):
  5, 10 and 20. The mock spectra are constructed from a same intrinsic 
  dust-free spectrum, and are reddened with a Calzetti attenuation curve
  assuming a color excess of $E(B-V)=0.2$. The top panels display the
  mock spectra, and panels in the second row show the small-scale and
  large-scale components of each spectrum. In the third row, the black 
  line in each panel is the ratio of small-scale to large-scale component
  (S/L), while and the red line plots the S/L of the best-fitting model spectrum. 
  In the fourth row, the black line is the recovered relative attenuation 
  curve and the red line is the input attenuation curve. 
  The $E(B-V)$ estimated from the measured attenuation curves is 0.221, 
  0.194, and 0.205 respectively, as indicated. In the bottom panels, the
  black lines show the ratios of the measured attenuation curve (smoothed
  for clarity) to the input one. The root square mean error (RMSE) of these
  curves is 2.9\%, 1.7\% and 0.8\%, also indicated.}
  \label{fig:example}
\end{figure*}

\subsection{Examples} \label{subsec:spec_exa}

Figure~\ref{fig:example} shows some examples to demonstrate
the step-by-step application of our method to measure the relative dust 
attenuation curve, $A(\lambda)-A_V$, as well as the color excess, $E(B-V)$. 
The three panels in the first row display three 
mock spectra that correspond to $E(B-V)=0.2$ 
but have different Gaussian signal-to-noise ratios 
SNR=5, 10, and 20, as indicated on the top of the figure.  
Here a Calzetti dust curve \citep{2000ApJ...533..682C} 
is adopted to create the mock spectra, and the procedure to construct 
the mock spectra will be detailed in the next section. In  
the second row, the moving average filter (see Eqn.~\ref{eq:large} 
and~\ref{eq:small}) is applied to decompose each spectrum 
into the large-scale and small-scale components, which
are plotted separately in each panel. The third row shows 
the small- to large-scale spectral ratio in black lines,
and the best-fitting ratio (see Eqn.~\ref{eq:fit1})
in red lines, as obtained from the non-linear fitting described above. 
The coefficients obtained from the fitting are then used to 
derive the best-fitting model spectrum (see Eqn.~\ref{eq:fit2}), 
which is expected to be dust-free. The relative dust attenuation 
curves, given by the ratio between the original spectra
and the best-fitting models (see Eqn.~\ref{eq:fit5}), are shown 
as the black lines in the fourth row. The input attenuation curve
is repeated in the three panels as a red line for comparison. 
The $E(B-V)$ values obtained from our 
measured attenuation curves (see Eqn.~\ref{eq:ebv}) are 0.221, 
0.194 and 0.205 for the three SNRs, respectively, 
as indicated in each panel, all very close to the input value, 
$E(B-V)=0.2$.
In the bottom panels, we show the ratios of the measured attenuation
curve to the input attenuation curve, which are smoothed for clarity. 
As one can see, our method recovers the input attenuation curve very well. 
The rms deviations of the measured attenuation curves around the 
input one are $2.9\%$, $1.7\%$ and $0.8\%$ for the three 
SNRs, respectively.

\begin{figure}
    \includegraphics[width=0.45\textwidth]{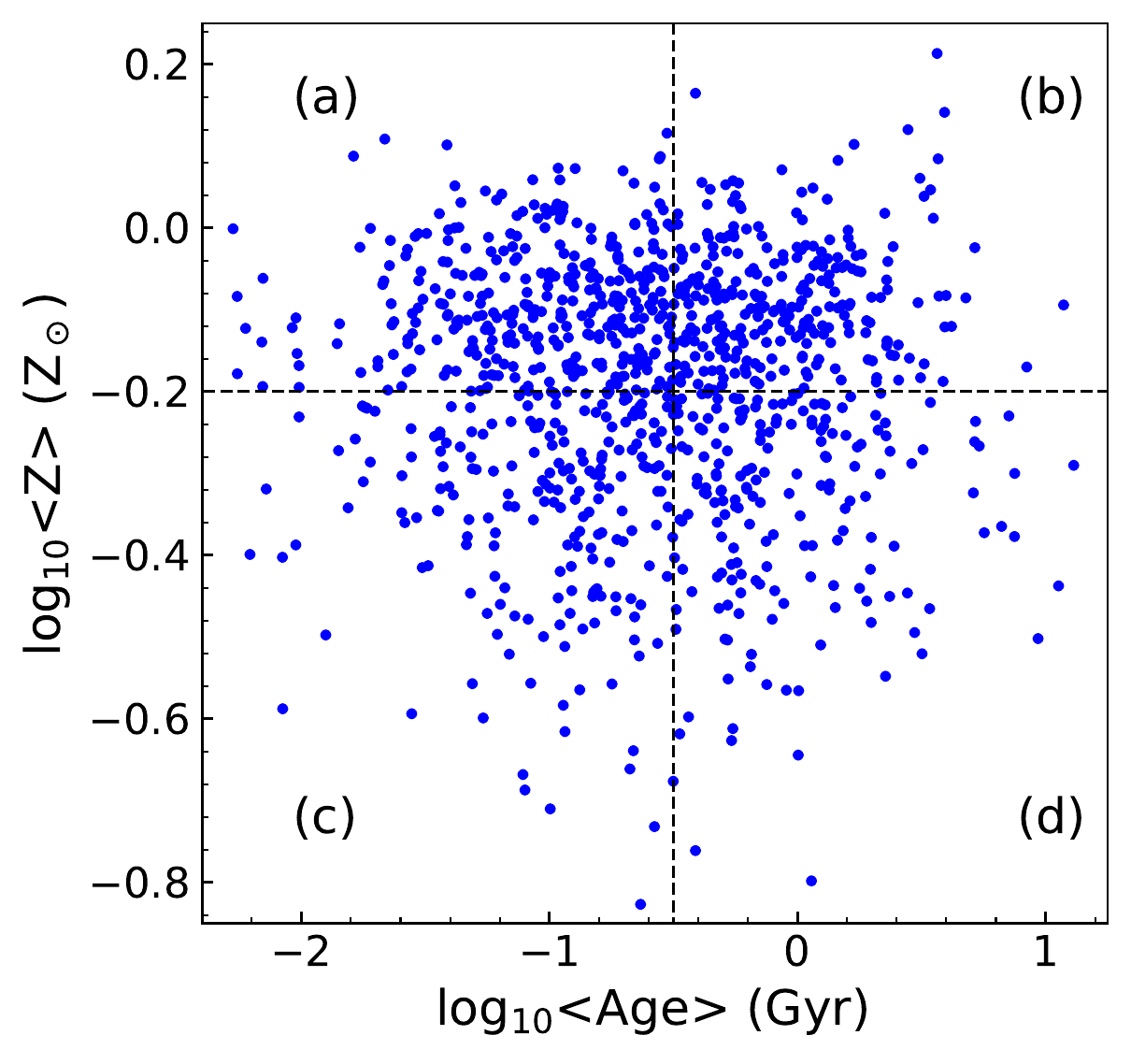}
    \caption{The distribution of mean age and mean metallicity of the mock spectra. 
    We divide the age-metallicity space roughly into four regions,  
    (a), (b), (c) and (d).}
    \label{fig:agez}
\end{figure}

\begin{figure}
    \includegraphics[width=0.45\textwidth]{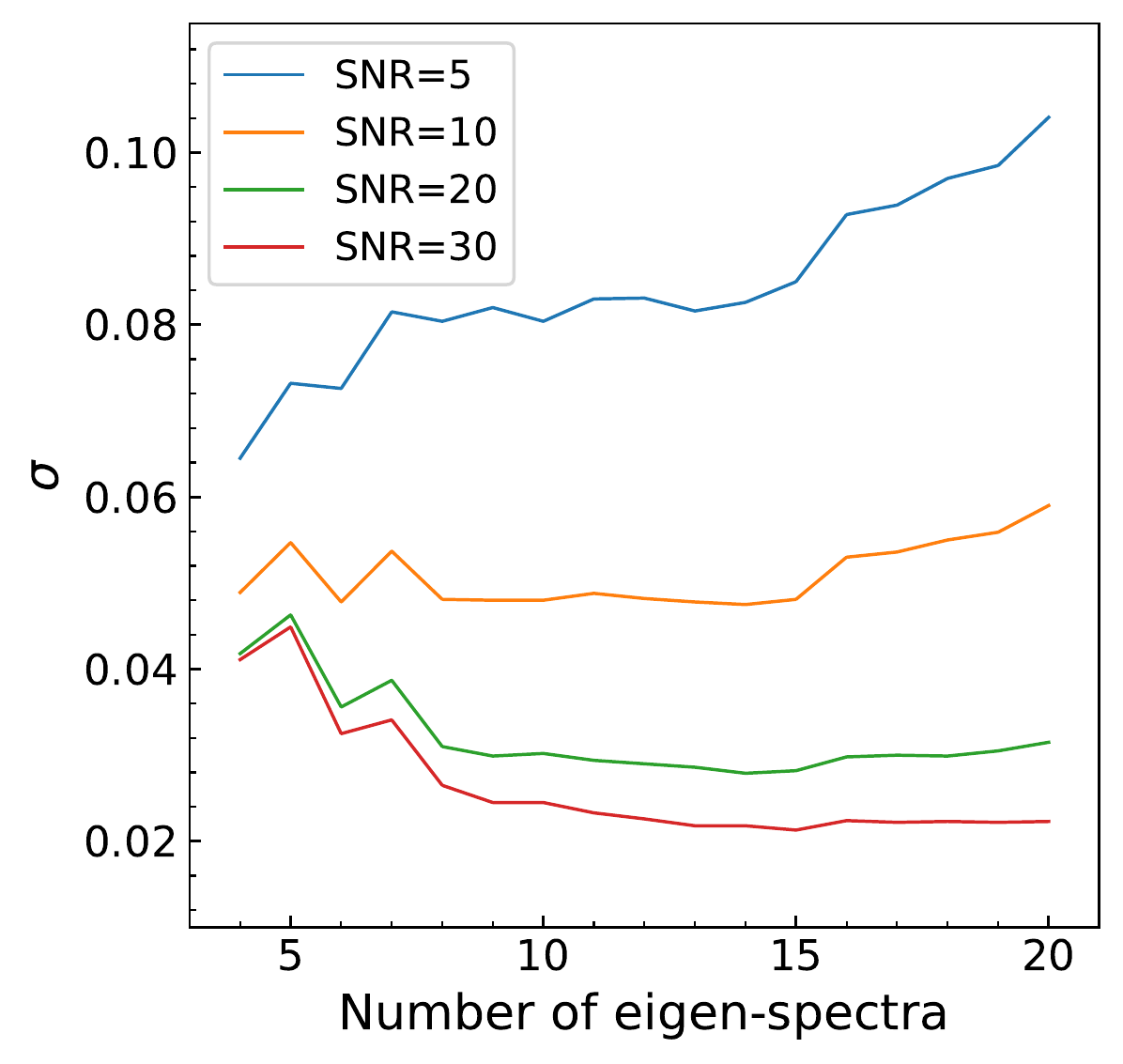}
    \caption{Tests on the effects of different number of eigen-spectra.
     Here the standard deviation ($\sigma$) of the difference 
     between the output and input $E(B-V)$ values is plotted 
     against the number of eigen-spectra. SNRs are indicated in this figure.}
    \label{fig:num_eigen}
\end{figure}

\begin{figure*}
  \centering
    \includegraphics[width=1\textwidth]{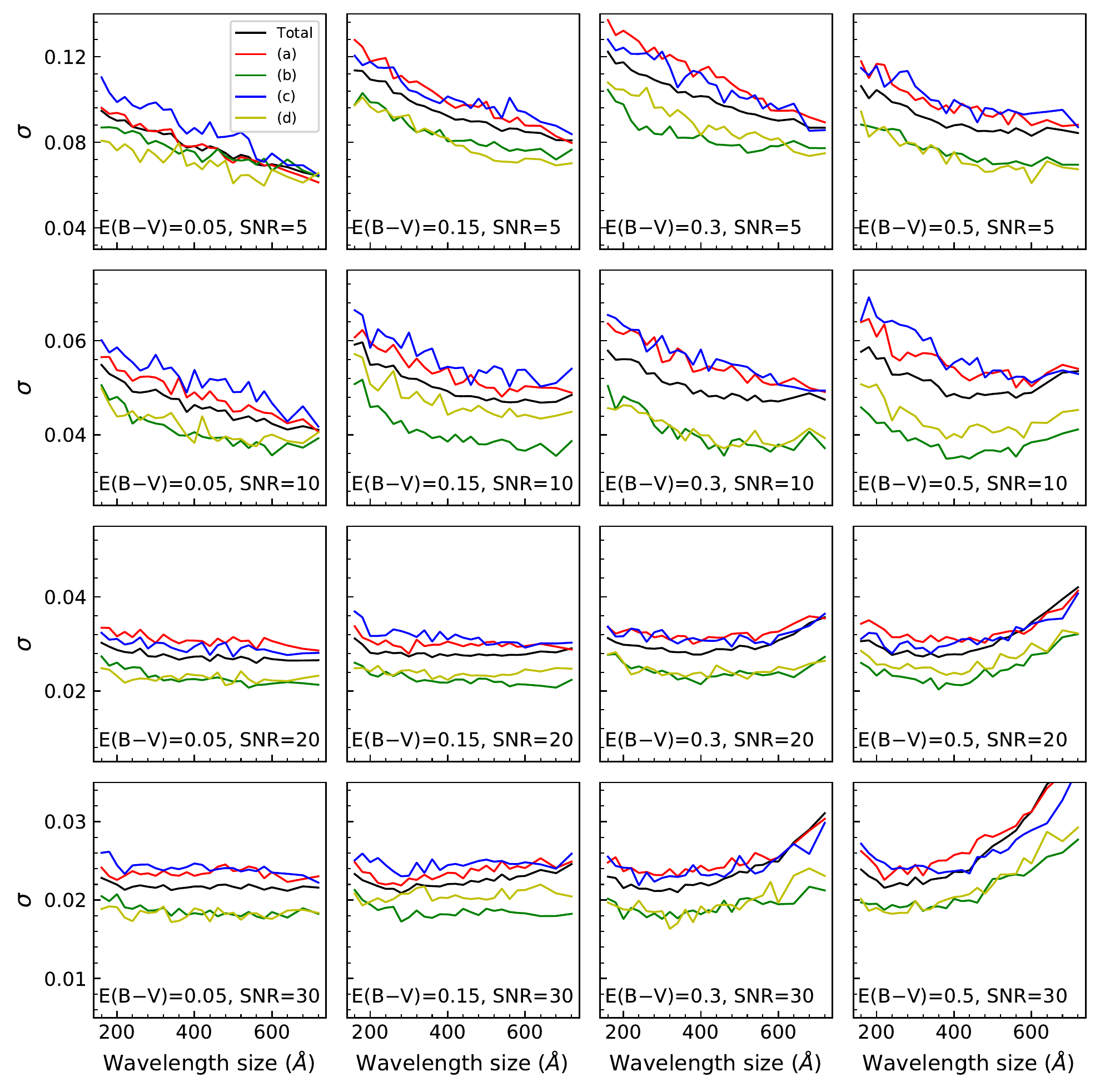}
    \caption{Tests on the effects of the wavelength window size 
     used to define the small- and large-scale components of spectra.
     Here the standard deviation ($\sigma$) of the difference 
     between the output and input $E(B-V)$ values is plotted 
     against the wavelength window size. In each panel, the input 
     value of $E(B-V)$ and the signal-to-noise ratio, SNR  
     are indicated. The red, green, blue, yellow lines correspond 
     to spectra in the (a), (b), (c), (d) regions, respectively. 
     The black line is the result for all the spectra in the four 
     regions.}
    \label{fig:wavewin}
\end{figure*}

\section{Test with mock spectra} \label{sec:mock}

\subsection{The mock spectra} \label{subsec:mock_spec}

In order to test the reliability of our method, we have created 
a series of synthetic spectra from the BC03 SSPs. We first pick 150 SSPs
out of the 1326 BC03 SSPs, with 25 different ages and 6 different metallicities. 
For each metallicity, the 25 SSPs are chosen so as to cover the full 
range of age from 0.001 Gyr to 18 Gyr, with approximately equal
intervals in the logarithm of the age. We then randomly select
one of the 25 SSPs for every metallicity, and the six selected spectra 
of different metallicites and ages are then normalized at 5500\AA\ 
and combined with random weights. We repeat this step 1000 times, 
thus creating 1000 synthetic mock spectra with a wide coverage of 
age and metallicity, as shown in Fig.~\ref{fig:agez}. 
In the figure, the age-metallicity space is roughly divided into 
four regions as indicated by the vertical and horizontal lines:
(a) young age and high metallicity, (b) old age and high metallicity, 
(c) young age and low metallicity, (d) old age and low metallicity.
Each of the mock spectra is reddened with a Calzetti dust curve 
\citep{2000ApJ...533..682C}, but assuming four different color
excesses: $E(B-V)=0.05$, $0.15$, $0.3$, and $0.5$. Finally, a 
Gaussian noise with SNR=5, 10, 20, or 30 is added to each spectrum. 
This procedure results in a total of 16,000 mock spectra, which 
are used to test our method. 

We would like to point out that some of the mock spectra may not be  
physically meaningful, e.g. those with extremely old ages but high
metallicities and those with young ages but low metallicities.
We do not exclude these spectra from our test, 
because the inclusion of them does not affect our test results,
as we will show below.

\subsection{The choice of eigen-spectra number}

As mentioned, we adopt the first 14 eigen-spectra
resulted from the PCA analysis as the model templates for
our spectral fitting. With the mock spectra generated above,
we have examined the potential effect of using different numbers 
of eigen-spectra on the measurement of $E(B-V)$.
For a given number of eigen-spectra and the mock spectra
with a given SNR, we calculate the standard deviation, 
$\sigma$, of the difference between the output and input $E(B-V)$
values, and we use this quantity to indicate the 
goodness of our method in recovering the true $E(B-V)$. 
The result of this analysis is shown in 
Fig.~\ref{fig:num_eigen}, which plots $\sigma$ as function 
of the number of eigen-spectra for four different SNRs.

The figure shows that, for mock spectra with SNR$>10$, 
the value of $\sigma$ decreases as one adopts more and more 
eigen-spectra in the fitting, but $\sigma$ does not decrease
any more when the number of eigen-spectra exceeds $\sim15$.
The value of $\sigma$ keeps roughly constant for SNR=10
and even increases for SNR=5. In spectra with low SNRs, 
features on small scales such as stellar absorption lines 
can be well dominated by noise, and so the use of too many 
eigen-spectra actually leads to poor fits. Overall, the figure 
indicates that the use of $8$-$15$ eigen-spectra appears to 
be able to achieve a compromise among different SNRs. 
We opt for the first 14 eigen-spectra considering mainly
the fact that the $\sigma$ starts to increase with 15
eigen-spectra at all SNRs. Nevertheless, we note that 
the variation of $\sigma$ with the number of eigen-spectra
used is in general modest in comparison to that produced 
by different SNRs.

\subsection{The choice of filtering window size} \label{sec:windowsize}

As mentioned before, filtering to separate a spectrum into 
small- and large-scale components plays an important role in 
our method, and it is critical to find a suitable choice of 
the wavelength window size for the filtering. For this purpose, 
we have adopted window sizes ranging from 160\AA\ to 720\AA\
when filtering the mock spectra. Fig.~\ref{fig:wavewin} shows 
$\sigma$ versus the wavelength window size, for mock spectra 
of different $E(B-V)$ and SNRs. Defined in the same way as above,
the $\sigma$ is an indicator of the goodness of our method.
The four colorful curves in each panel correspond to the four 
regions divided in Fig.~\ref{fig:agez}, while the black curve
is for all the spectra with the given $E(B-V)$ and SNR. 
Generally, as expected, the value of $\sigma$ decreases as SNR 
increases. In addition, for low values of SNR (5 and 10), 
$\sigma$ tends to decrease as the wavelength window size 
increases, but the trend reverses when SNR and $E(B-V)$ are high.

The results shown in Fig.~\ref{fig:wavewin} suggest that 
the wavelength window size should be chosen to be 
about 500\AA\ for SNR smaller than 10 and about 300\AA\ for 
higher SNR. This dependence of smoothing window size on SNR is 
understandable. A larger window size can reduce the effects 
of noise in the low SNR spectra more effectively, while a 
smaller window size is needed to retain more real features 
in the spectra with high SNR. Spectra with older stellar ages
(green and yellow lines) also trend to have lower $\sigma$ 
than younger ones (red and blue lines), indicating
that our method works better for older stellar populations.
However, the differences in the results among the four 
different regions are not large in comparison with the effects 
of the SNR and the wavelength window size.

\begin{figure*}
  \centering
    \includegraphics[width=1\textwidth]{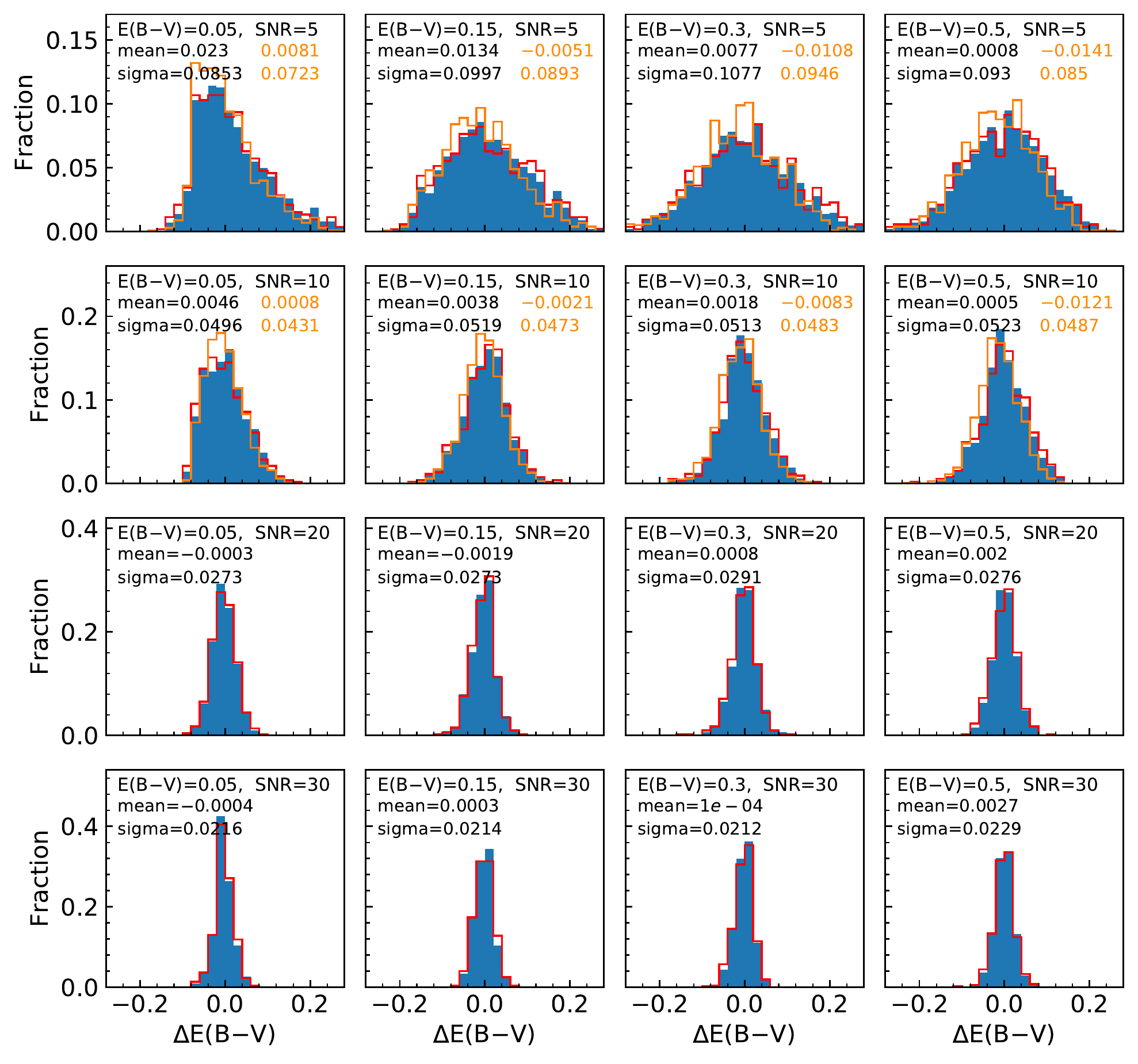}
    \caption{Tests with mock spectra that have different $E(B-V)$ values and 
    signal-to-noise ratios (SNR). 
    Here the distribution is shown with respect to $\Delta E(B-V)$), 
    the difference between the fitting result of $E(B-V)$ and the input value. 
    The input $E(B-V)$ and SNR are indicated in each panel.
    The mean value and standard deviation of $\Delta E(B-V)$ are indicated 
    by `mean' and `sigma', respectively. For all cases, we choose a wavelength window size of 300\AA. The red histograms show the results of `unreasonable' spectra that locate in upper-right and lower-left corners of Fig.~\ref{fig:agez}. For lower-SNR spectra with SNR=5 and 10, we also show the test for a window size of 500\AA, plotted as the orange histograms in the top two rows.}
    \label{fig:mock}
\end{figure*}

\begin{figure*}
  \centering
    \includegraphics[width=1\textwidth]{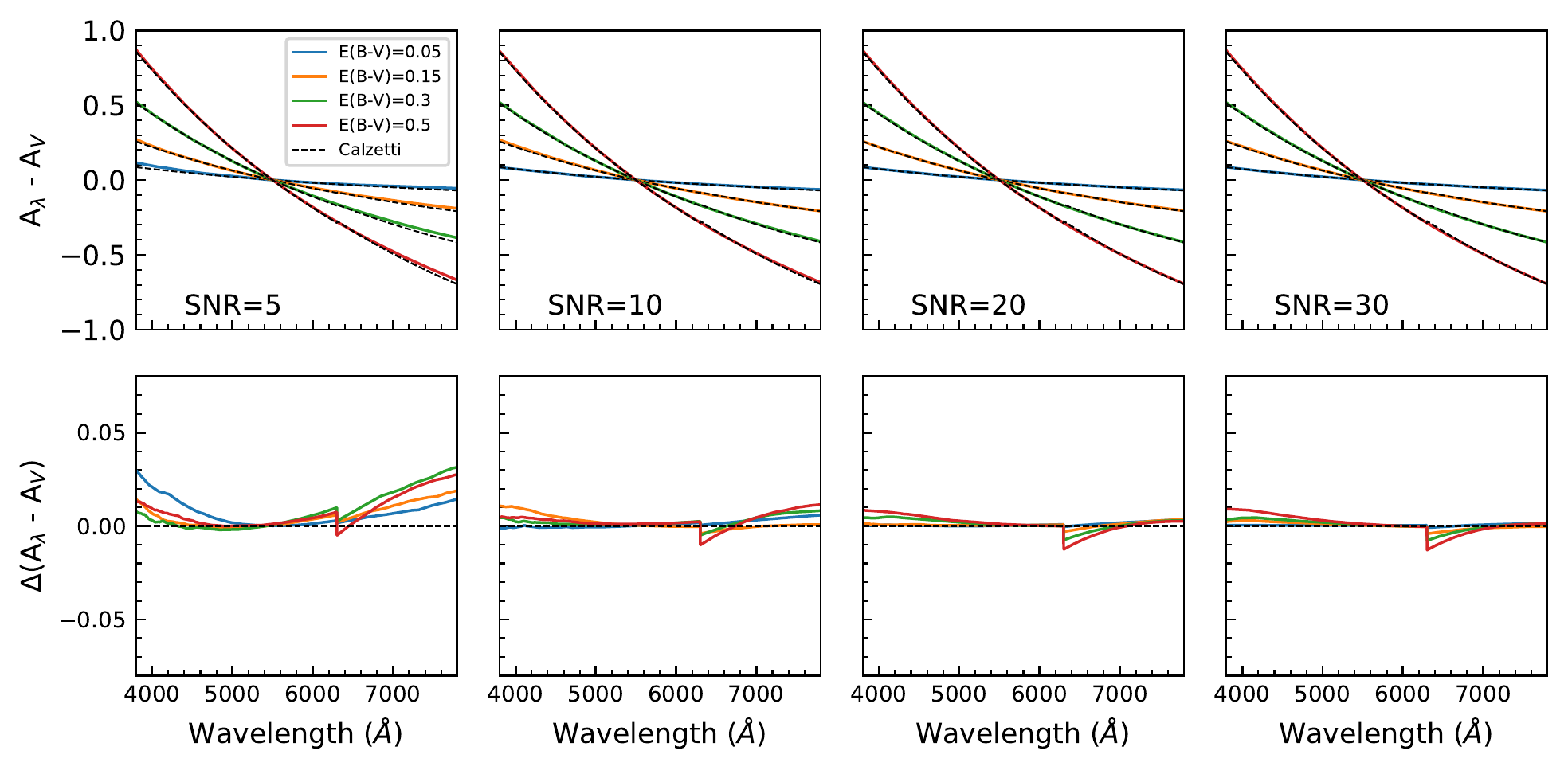}
    \caption{The first row shows the medians of all relative dust attenuation curves in each $E(B-V)$ and SNR bins.
    The second row shows the differences between the median curves and the Calzetti curve.}
    \label{fig:mock_curve}
\end{figure*}

\subsection{Test results}

Fig.~\ref{fig:mock} shows the results of our test on mock spectra.
Each panel is based on 1000 spectra with given $E(B-V)$ and SNR, 
as described in \S\ref{subsec:mock_spec}. The filled histogram in each
panel shows the fraction of the 1000 fitting as a function of  
the difference between the output and input values of $E(B-V)$, 
referred to as $\Delta E(B-V)$ in what follows.
In each panel, the input values of $E(B-V)$ and SNR, the mean
and standard deviation ($\sigma$) of $\Delta E(B-V)$, are indicated.
We have chosen a wavelength window size of 300\AA\ for all the 
cases, according to the analysis presented in the previous subsection. 
For low-SNR spectra with SNR=5 and 10, we repeat the test 
using a larger window size of 500\AA, and plot the results as 
the orange histograms in the top two rows. As can be 
seen, the results change little with varying wavelength window sizes.

The standard deviation, $\sigma$, decreases as SNR increases,  
but its dependence on the input $E(B-V)$ is quite weak.  
The distribution of $\Delta E(B-V)$ in most cases is roughly a Gaussian 
centered at $\Delta E(B-V)\sim 0$, indicating that the model inference 
is unbiased. The only exception is the case shown in the first panel,
where the distribution is skewed to positive values of $\Delta E(B-V)$.
In this case, the inputs of $E(B-V)$ and SNR are both small. 
We note that increasing the wavelength window size for the moving
average filter does not make a significant improvement of the $E(B-V)$ 
measurement in such low-SNR and low-$E(B-V)$ cases.

We also show the results for ``unreasonable'' mock 
spectra as the red histograms in individual panels. These spectra
are located in regions $b$ and $c$ in Fig.~\ref{fig:agez}:  
they either have old ages and high metallicities, or young ages and 
low metallicities. The distributions of $\Delta E(B-V)$ for these 
spectra are almost the same as those of the full samples 
at given $E(B-V)$ and SNR, indicating that the inclusion of such 
spectra does not introduce biases into our test results. 

As described in \S\ref{subsec:spec_fit}, our method can recover the 
relative attenuation curve. As a demonstration, the upper row of 
Fig.~\ref{fig:mock_curve} shows the median $A_\lambda-A_V$ curve 
at given $E(B-V)$ and SNR. Panels from left to right are for
different SNR bins, and the different colors in each panel are
for different $E(B-V)$ bins. For each case the corresponding
Calzetti curve is plotted as a dotted line. The differences between
the median curves output from our method and the input Calzetti 
curve are shown in the lower row. As one can see, our method 
indeed recovers the input curve very well in all cases and at all
wavelengths. The largest difference is $\sim0.03$ magnitude, 
and this occurs at lowest SNR (SNR=5) and at both short ($<5000$\AA)
and long wavelengths ($>6500$\AA). At higher SNRs the difference
is generally very small, less than 0.01 magnitude at all wavelengths.
There is a slight but noticeable jump at $\sim6200$\AA\ in all
cases. This is caused by the fact that the Calzetti curve is a
piecewise function with two sub-functions separated at a fixed
wavelength, while the attenuation curves output from our method
are smooth, thus not necessarily following any predefined functional
forms.

To conclude, the results of the test demonstrate that the input 
$E(B-V)$ are well recovered by our method for various cases. 
The systematic bias in $\Delta E(B-V)$ is quite small, indicating 
that potential uncertainties, such as that produced by 
the dust-age degeneracy, is well taken care of by our method.

\begin{figure*}
  \centering
  \fig{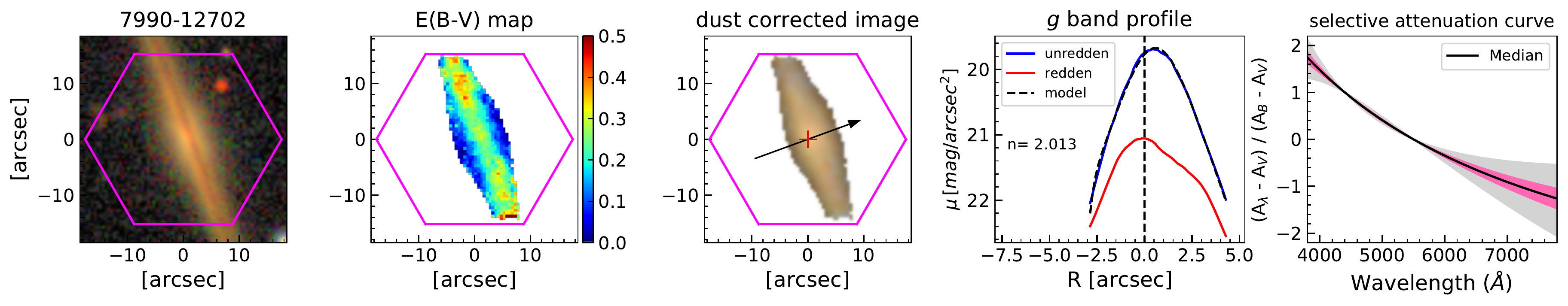}{1\textwidth}{}
  \vspace{-0.2 cm}
  \fig{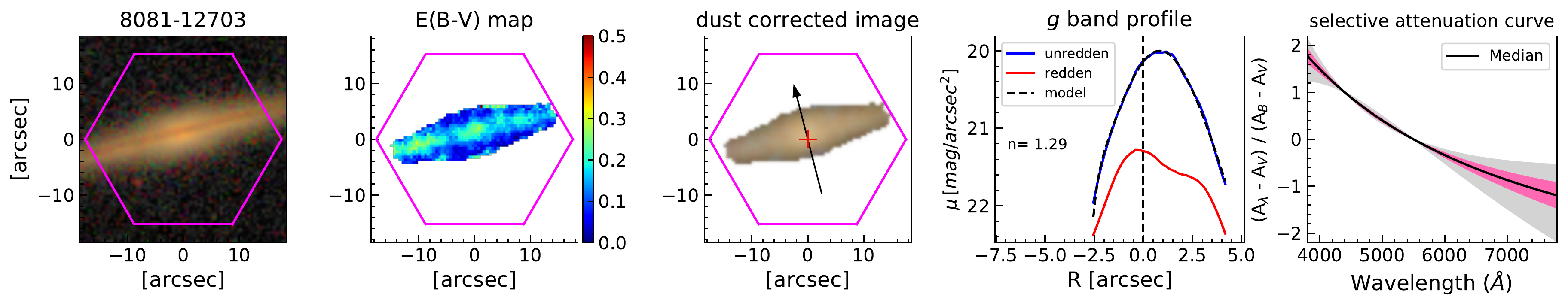}{1\textwidth}{}
  \vspace{-0.2 cm}
  \fig{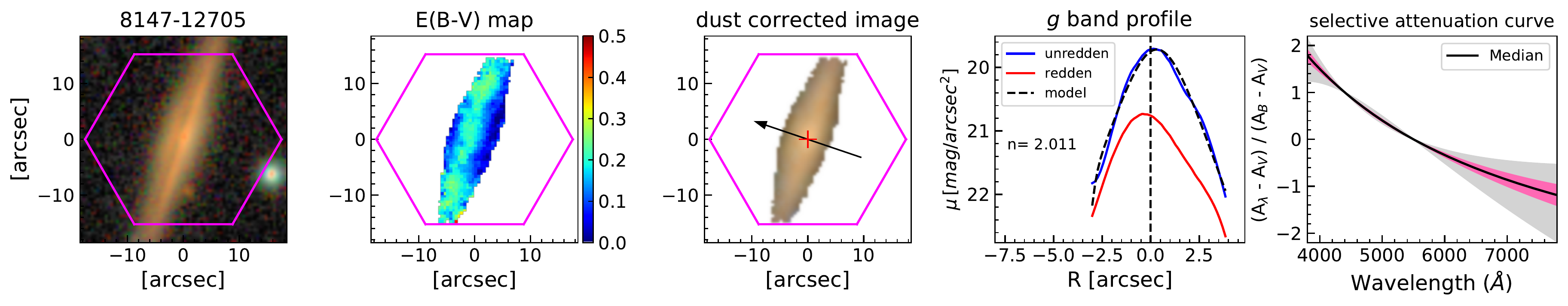}{1\textwidth}{}
  \vspace{-0.2 cm}
  \fig{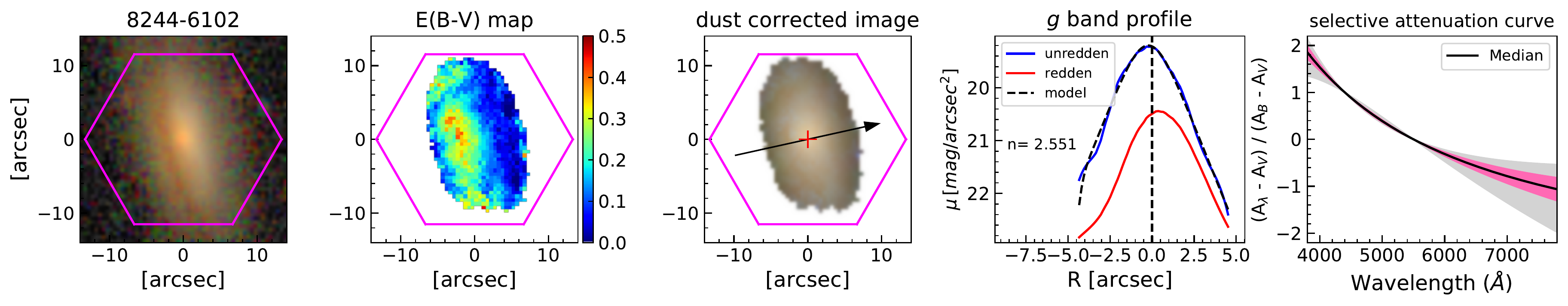}{1\textwidth}{}
  \vspace{-0.2 cm}
  \fig{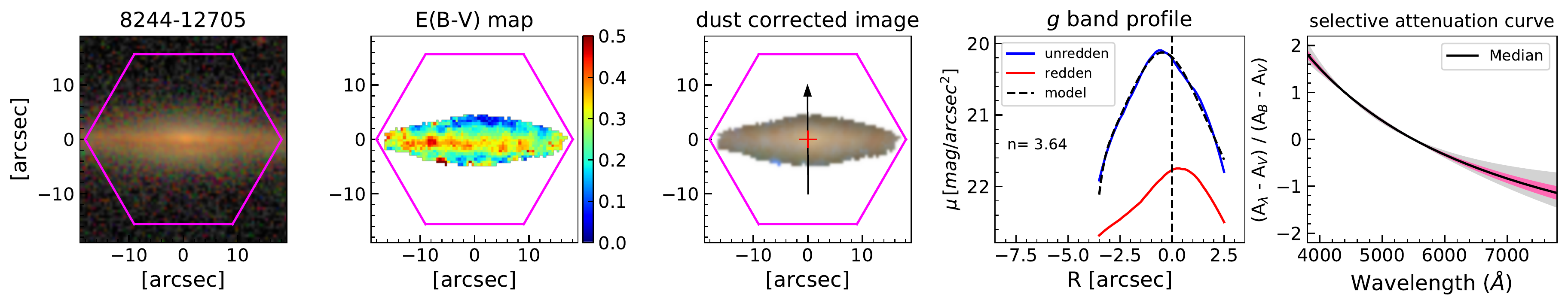}{1\textwidth}{}
  \caption{There are five panels in each row. The first one is SDSS $gri$ color-composite image with the MaNGA footprint in magenta. The second one is the $E(B-V)$ map derived from our method. The third one is dust corrected $gri$ color-composite image. The red cross shows the center of the integral field. In the fourth panel, the red line is reddened brightness profile while the blue line is un-reddened brightness profile along the direction of arrow in the third panel. The un-reddened brightness profile is fitted by a model that the black dash line shows. $n$ is the key parameter of this model. The last panel shows the selective attenuation curves measured from our method. The black solid line is the median, the pink region shows the standard deviation of the spaxels around the median, and the grey region covers the range spanned by all the individual spaxels.}
   \label{fig:manga}
\end{figure*}

\addtocounter{figure}{-1}
\begin{figure*}
  \centering
  \fig{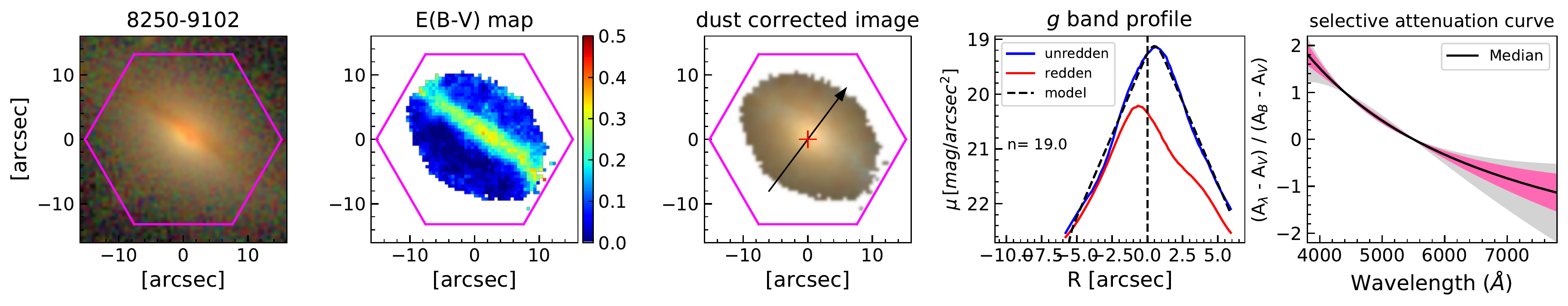}{1\textwidth}{}
  \vspace{-0.2 cm}
  \fig{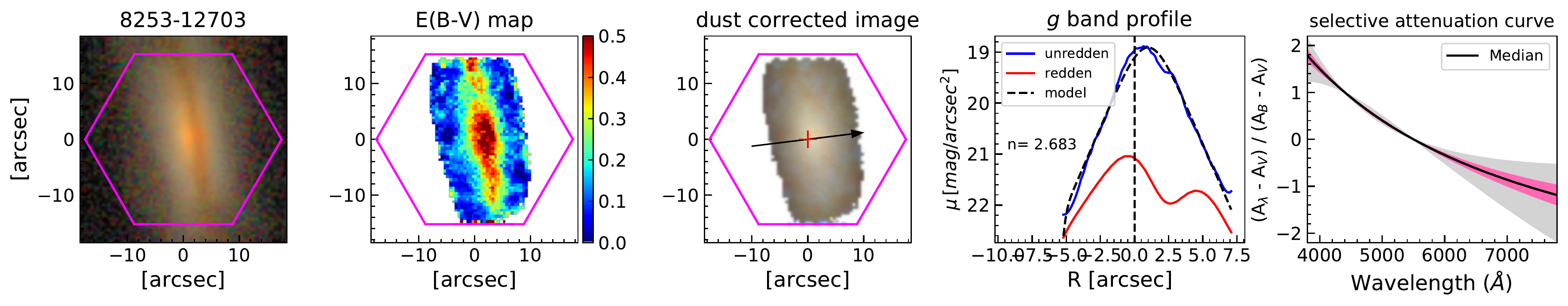}{1\textwidth}{}
  \vspace{-0.2 cm}
  \fig{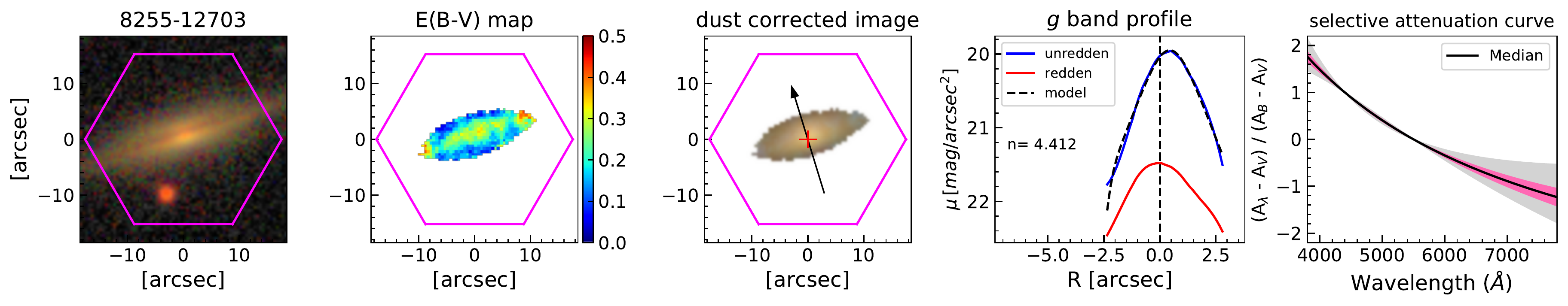}{1\textwidth}{}
  \vspace{-0.2 cm}
  \fig{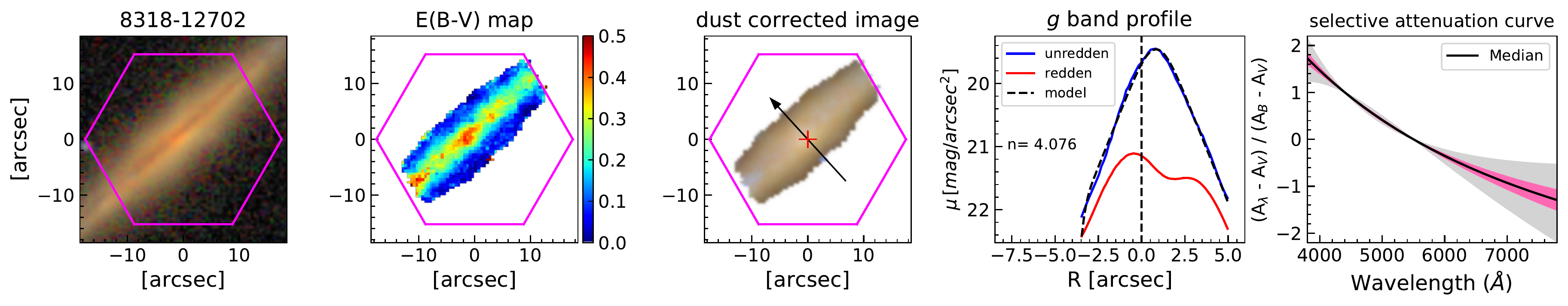}{1\textwidth}{}
  \vspace{-0.2 cm}
  \fig{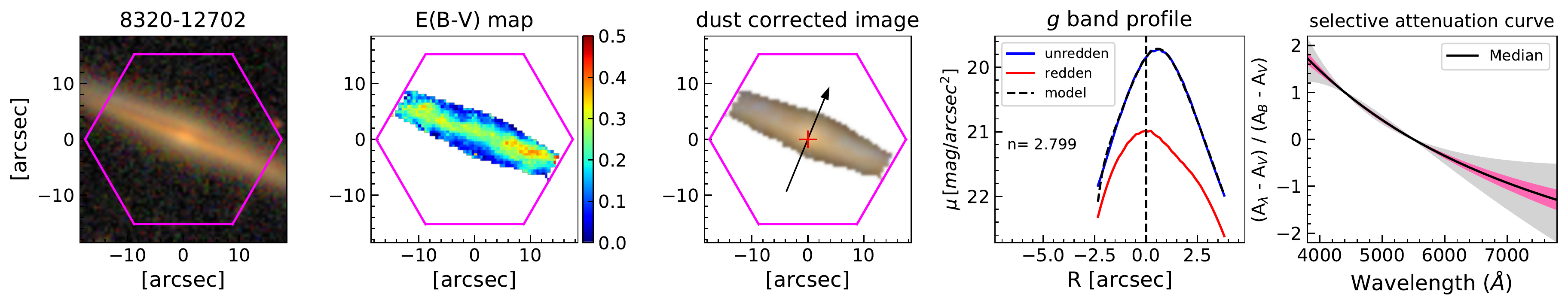}{1\textwidth}{}
  \caption{(Continued). There are five panels in each row. The first one is SDSS $gri$ color-composite image with the MaNGA footprint in magenta. The second one is the $E(B-V)$ map derived from our method. The third one is dust corrected $gri$ color-composite image. The red cross shows the center of the integral field. In the fourth panel, the red line is reddened brightness profile while the blue line is un-reddened brightness profile along the direction of arrow in the third panel. The un-reddened brightness profile is fitted by a model that the black dash line shows. $n$ is the key parameter of this model. The last panel shows the selective attenuation curves measured from our method. The black solid line is the median, the pink region shows the standard deviation of the spaxels around the median, and the grey region covers the range spanned by all the individual spaxels.}
\end{figure*}

\addtocounter{figure}{-1}
\begin{figure*}
  \centering
  \fig{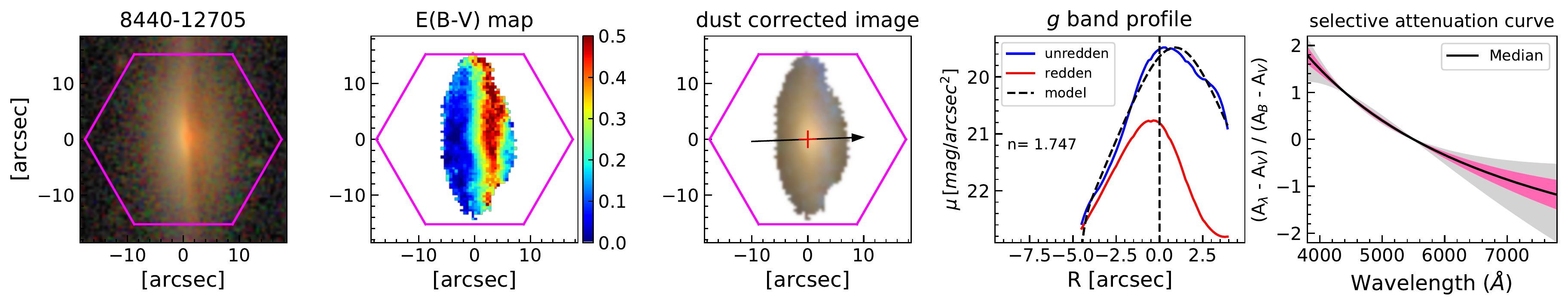}{1\textwidth}{}
  \vspace{-0.2 cm}
  \fig{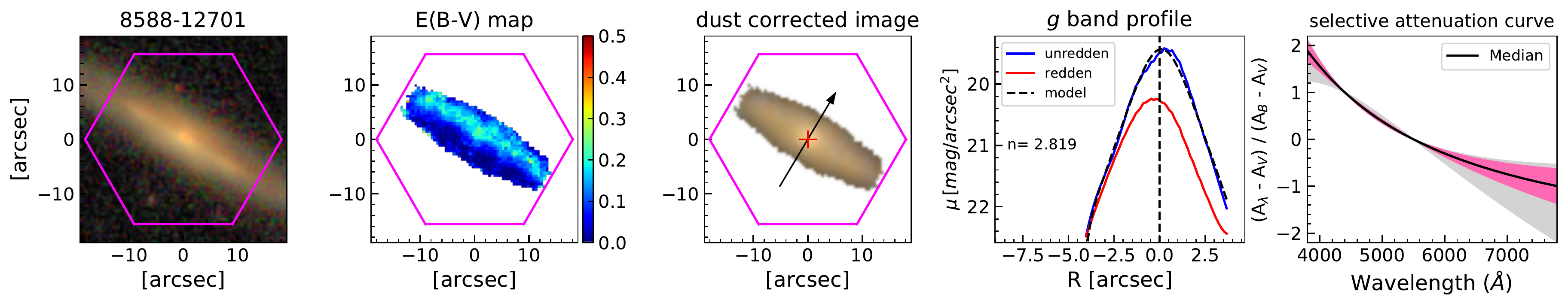}{1\textwidth}{}
  \vspace{-0.2 cm}
  \fig{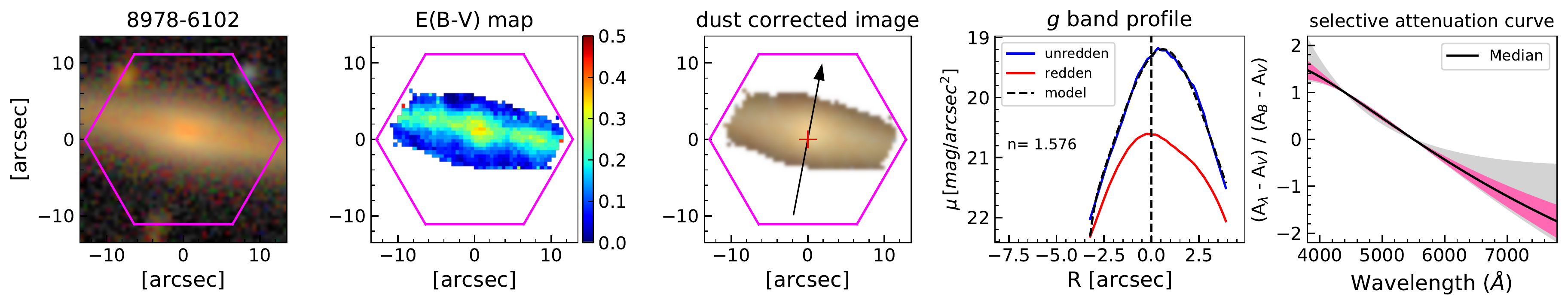}{1\textwidth}{}
  \vspace{-0.2 cm}
  \fig{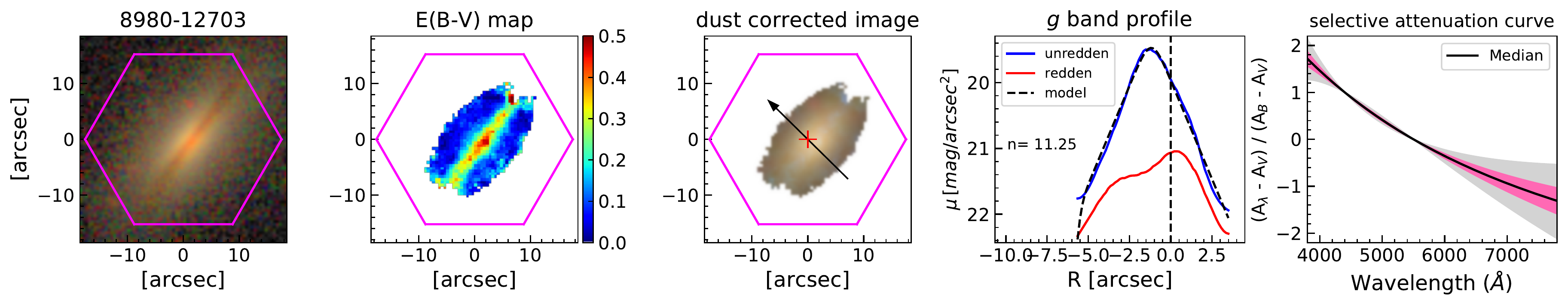}{1\textwidth}{}
  \vspace{-0.2 cm}
  \fig{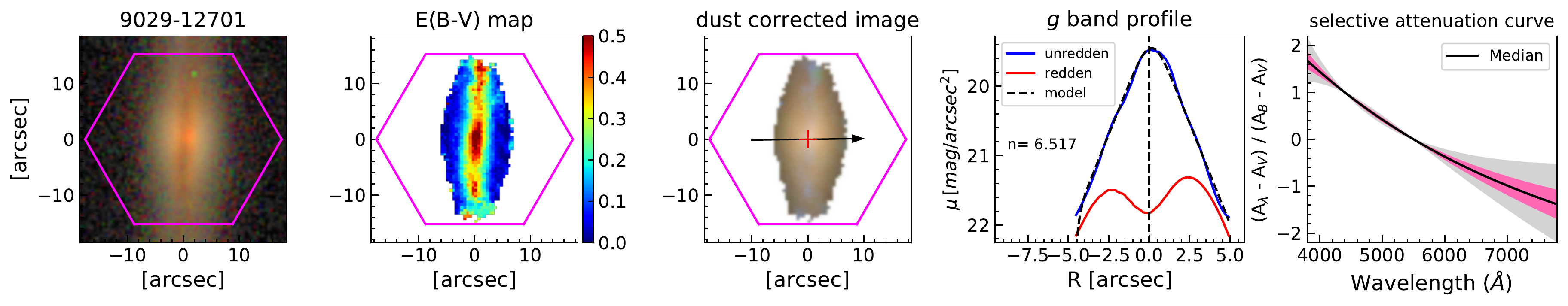}{1\textwidth}{}
  \caption{(Continued). There are five panels in each row. The first one is SDSS $gri$ color-composite image with the MaNGA footprint in magenta. The second one is the $E(B-V)$ map derived from our method. The third one is dust corrected $gri$ color-composite image. The red cross shows the center of the integral field. In the fourth panel, the red line is reddened brightness profile while the blue line is un-reddened brightness profile along the direction of arrow in the third panel. The un-reddened brightness profile is fitted by a model that the black dash line shows. $n$ is the key parameter of this model. The last panel shows the selective attenuation curves measured from our method. The black solid line is the median, the pink region shows the standard deviation of the spaxels around the median, and the grey region covers the range spanned by all the individual spaxels.}
\end{figure*}

\begin{figure}
  \centering
  \fig{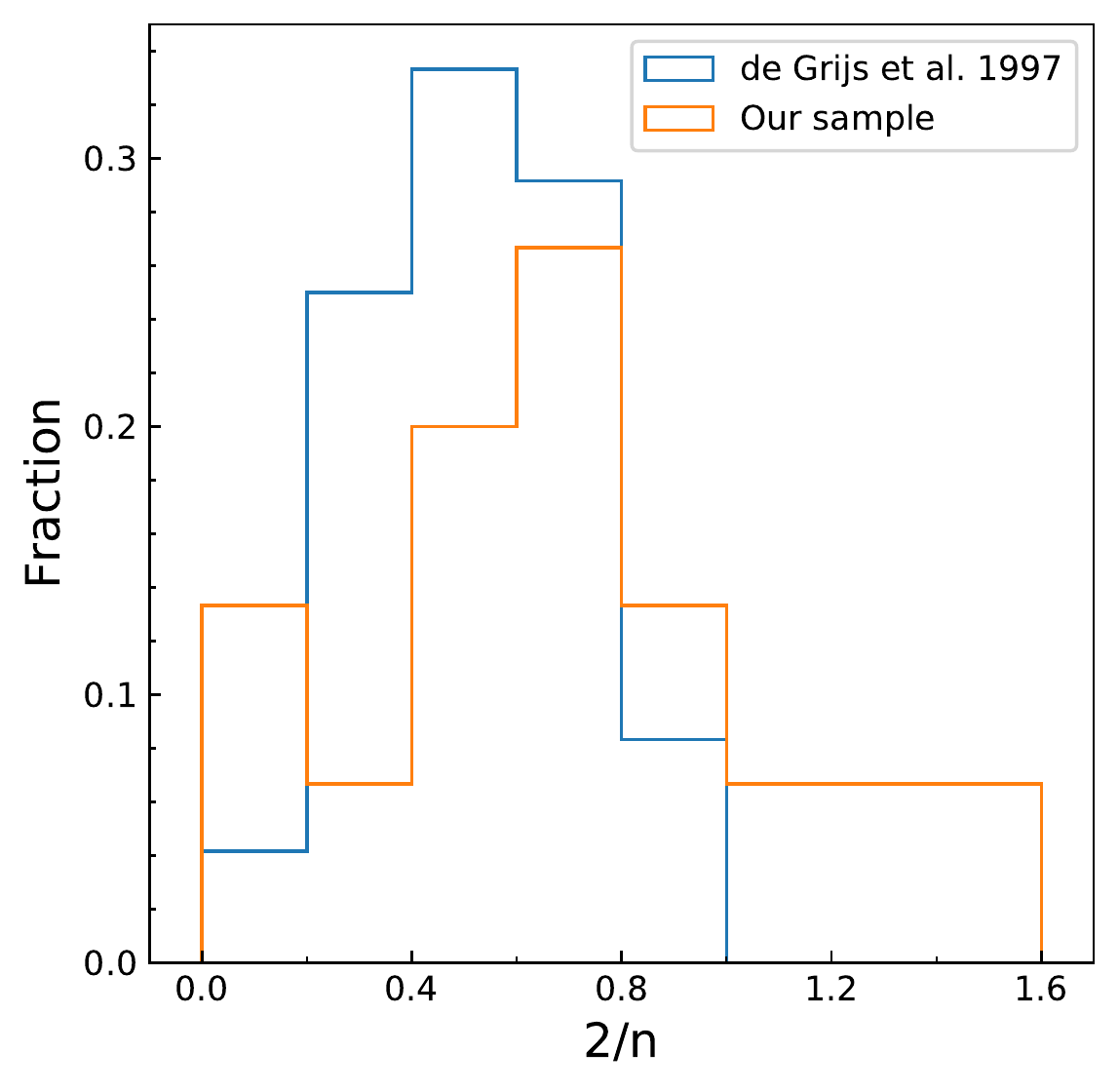}{0.45\textwidth}{}
  \caption{The distribution of parameter $2/n$, which is a parameter
  in \autoref{eq:fz} controlling the central sharpness of a vertical 
  surface brightness profile.
  The blue histogram is the result from our sample, while the 
  orange histogram is obtained from \citet{1997A&A...327..966D}.}
  \label{fig:n}
\end{figure}

\begin{figure*}
  \centering
    \fig{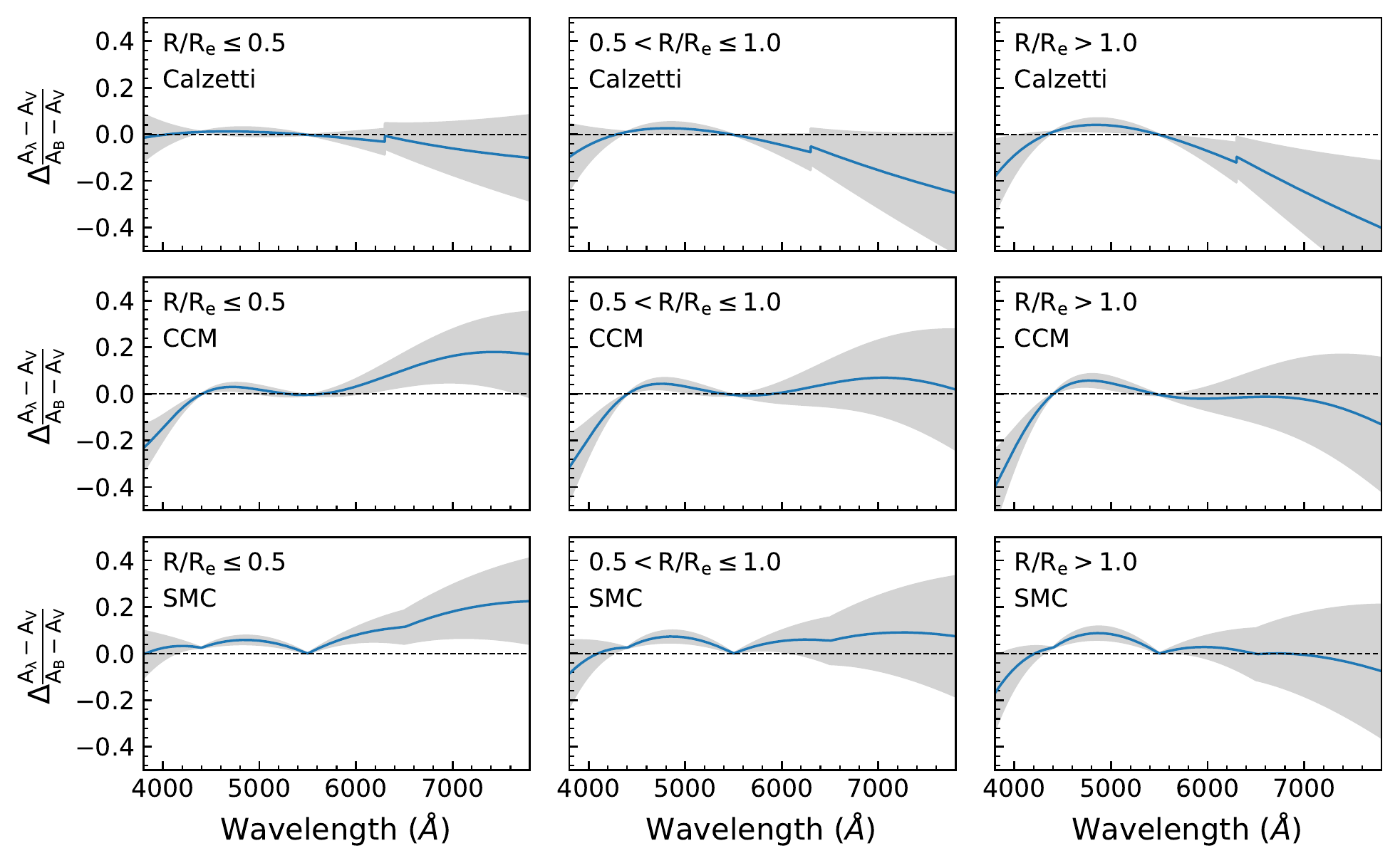}{\textwidth}{}
    \caption{The differences of the fitted Calzetti (upper panels), CCM (middle panels) and SMC (bottom panels) model curves relative to the measured selective attenuation curve of the 15 MaNGA galaxies. Panels from left to right are the results for spaxels at three different radial intervals as indicated. In each panel, the solid blue line is the median of all the spaxels at the given radii, and the grey region indicates the standard deviation of the spaxels.}
   \label{fig:curve_bins}
\end{figure*}

\begin{figure*}
  \centering
    \fig{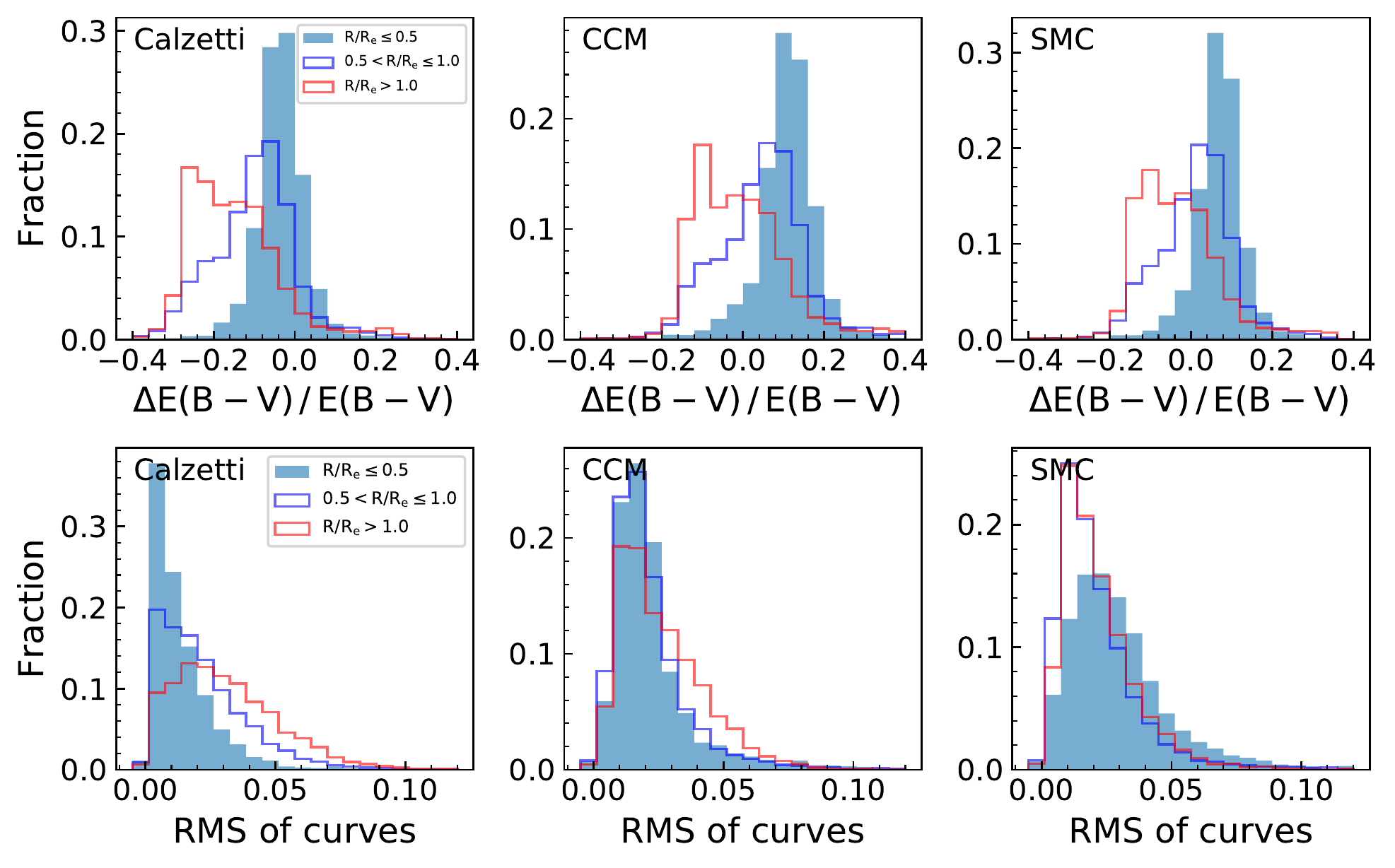}{\textwidth}{}
    \caption{{\it Upper:} Distribution of the relative difference of $E(B-V)$ between the fitted model curves of Calzetti, CCM and SMC (panels from left to right) and the measured curve. Plotted in different colors are results for spaxles at different radii, as indicated.
    {\it Lower:} Distribution of the rms scatter of the fitted curves around the measured one.}
   \label{fig:curve_rms}
\end{figure*}

\section{Applications to MaNGA galaxies} \label{sec:manga}
\subsection{The MaNGA data}\label{subsec:data}

Mapping Nearby Galaxies at Apache Point Observatory (MaNGA) survey 
\citep{2015ApJ...798....7B,2016AJ....151....8Y,2016AJ....152..197Y,
2016AJ....152...83L,2017AJ....154...86W} is one of the three core programs 
of SDSS-IV \citep[Sloan Digital Sky Survey IV,][]{2017AJ....154...28B}. 
MaNGA aims to obtain spatially resolved spectroscopy of about 
10,000 nearby galaxies. In the outskirts of galaxies, MaNGA 
can reach a SNR of 4--8 (per \AA\ per $2\arcsec$ fiber) at 23 AB 
mag arcsec$^{-2}$ in the $r$-band. The wavelength coverage is 
between 3,600 and 10,300\AA\, with a spectral resolution 
$R\sim 2000$ \citep{2015AJ....149...77D}.
In this paper, we select galaxy samples from the Sloan Digital Sky 
Survey Data Release 14 \footnote{\url{http://www.sdss.org/dr14/manga/}} 
\citep[SDSS DR14,][]{2018ApJS..235...42A}, which includes integral
field spectroscopy (IFS) data from MaNGA for 2812 galaxies, 
including ancillary targets and $\sim 50$ repeated observations. 
We visually inspect the SDSS $gri$ color-composite images of 
all the galaxies in this release, and select 15 edge-on galaxies that 
have obvious dust lane features. Galaxies of this kind are much more 
affected by dust attenuation than those viewed from face-on, and thus 
provide more demanding tests of our method. In the rest of this paper, 
we will apply our method to the 15 galaxies to demonstrate that our 
method can effectively measure the spatially-resolved $E(B-V)$ maps and 
relative dust attenuation curves, even for these difficult cases.

\subsection{Dust attenuation maps and profiles}

In the current version of our code, velocity dispersion 
of the observed spectrum needs to be specified 
so that model templates can be adapted to the observed 
spectral resolution. We first use \texttt{PPXF}
\citep{2004PASP..116..138C,2017MNRAS.466..798C} to measure stellar 
kinematics and correct the broadening effect of velocity dispersion. 
We then apply our method to all the spatial pixels in the MaNGA
datacubes whose continuum SNRs are greater than 5. Fig.~\ref{fig:manga} 
displays the two-dimensional maps obtained from our method for the 15 
galaxies in our sample. The first panel in each row 
shows the SDSS $gri$ color-composite image of the galaxy in question, 
and the second panel is the $E(B-V)$ map obtained from 
our method. As one can see, the dust lanes in the original images
are well reproduced in the $E(B-V)$ maps, in terms of both the
spatial distribution and the relative strength. 

Once the $E(B-V)$ map is obtained for a galaxy, we can 
correct the effect of dust attenuation on the $gri$ image. 
Since the FWHM (full width at half maximum) of  
the point-spread function (PSF) for MaNGA is $\sim 2.5\arcsec$, 
while that of the SDSS $gri$ image is $\sim 1.4\arcsec$, 
we convolve the SDSS image in $g$, $r$ and $i$ bands with 
a Gaussian kernel to match the MaNGA PSF, before correcting 
the dust attenuation in the images.
As what we measure are relative, not absolute attenuation curves,
we assume a Calzetti attenuation curve to estimate the absolute
attenuation in $g$, $r$ and $i$ bands according to the estimated $E(B-V)$ map.
The images in the three bands are then corrected for the dust attenuation,
and combined to form a new $gri$ color-composite image,
which is expected to be dust free. This corrected image is shown
in the third panel for each galaxy. The dust lanes in the
original images are no longer seen in the dust-corrected
images. This can be seen more clearly in the fourth panel of each row, 
where the red and blue curves show, respectively, the original and 
corrected brightness profiles in the $g$-band, as measured along the 
arrow indicated in the third panel. Compared to the corrected brightness 
profile, the original one is not only lower in amplitude, but also 
asymmetric in shape due to the strong attenuation in dust lane regions.
In almost all cases, the prominent features in the vertical 
surface brightness profiles produced by the dust absorption 
are no longer present in the dust-corrected profiles. The corrected 
profiles are roughly symmetric with respect to the peak, as 
is expected for axis-symmetric thin disks. 

We attempt to fit the dust corrected brightness profiles
with a vertical brightness profile model.
Since galaxy disks are not infinitesimally thin, 
the three-dimensional luminosity density of the disk is typically written as
\begin{equation}
 \nu(R, z) = \nu_0 \exp(-R/R_d)f(z).
\end{equation}
The first part of this equation describes the surface brightness as a 
function of the radius $R$, which is assumed to have an exponential profile, 
with $R_d$ being the scale-length of the disk. The function 
$f(z)$ describes the surface brightness distribution in the vertical 
($z$) direction (see \S2.3.3 of \citet{2010gfe..book.....M} for 
a review). A commonly adopted fitting function of $f(z)$ is
\begin{equation}\label{eq:fz}
f_n(z)={\rm sech}^{2/n}\left( \frac{n|z|}{2z_d} \right),
\end{equation}
where $n$ is a parameter controlling the shape of profile near $z=0$,
and $z_d$ is the scale-height of the disk. In particular, 
$n=1$ corresponds to a self-gravitating isothermal sheet 
while $n=\infty$ corresponds to an exponential profile. 
Note that $f_n(z)$ decreases exponentially at large $|z|$,  
but a larger value of $n$ gives a steeper profile near the mid-plane.
When fitting the observed profile, the effect of seeing on 
the observed data must be included. This is done by 
convolving the model profile with a Gaussian kernel that
matches the MaNGA PSF. The best-fit model 
profile is plotted for each galaxy as the black dash line in the 
fourth panel of Fig.~\ref{fig:manga}. For reference, we also indicate 
the values of $n$ in the corresponding panels. It is clear that 
the corrected profiles are well described by the model. 

Fig.~\ref{fig:n} shows the distribution of $2/n$ for the 15 galaxies 
in comparison with that obtained by \citet{1997A&A...327..966D}
from fitting the $K$-band images of 24 nearly edge-on spiral galaxies. 
The two distributions are qualitatively consistent with each other, suggesting 
again that our method is able to correct dust attenuation for nearly 
edge-on galaxies. Unfortunately, no quantitative comparison can be made, 
as the two samples are small and have different selections.

\subsection{Dust attenuation curves}

As described in \S\ref{subsec:spec_fit}, one important advantage of 
our method is that the relative dust attenuation as a function of 
wavelength (i.e. the relative attenuation curve) can be directly
obtained without the assumption of a functional form for the shape
of the attenuation curve or the adoption of a theoretical dust model. 
The selective attenuation curves measured from our method are plotted 
in the rightmost panel of each row in Fig.~\ref{fig:manga}.
The grey region covers the range spanned by all the individual 
spaxels in a given galaxy. The black solid line is the median, 
while the pink region shows the standard deviation of the spaxels 
around the median.
The variance represented by the grey and pink regions 
  shows that
  different spaxels may have different shapes of attenuation curves,
  suggesting that dust attenuation may not be described completely 
  by a universal curve. Note that, by definition, the selective attenuation
  curves have a fixed value of unity at $\lambda_B$ (4400\AA) and zero at
  $\lambda_V$ (5500\AA), and this is why the variation among 
  individual spaxels appears only at  wavelengths shorter than $\sim
  4000$\AA\ and longer than $\sim6000$\AA.

In order to better understand the variation of the
  attenuation curve,  we further examine the measured selective
  attenuation curves at different radii from the galactic center. We
  also compare the measured curves with the
  commonly-adopted dust curves in the literature. Here we consider
  three dust curves: the Calzetti attenuation curve for local
  starburst galaxies \citep{2000ApJ...533..682C}, the Milky Way
  extinction curve \citep[][CCM]{1989ApJ...345..245C}, and the Small
  Magellanic Cloud extinction curve
  \citep[][SMC]{2003ApJ...594..279G}. We fit each of the measured
  attenuation curve with the three dust curves. 
  Fig.~\ref{fig:curve_bins} shows the differences  of the fitted
  selective attenuation curve with respect to the measured curve, for
  spaxels in the following three radial intervals: $R/R_e\le 0.5$, 
  $0.5 < R/R_e \le 1$ and
  $R/R_e>1$. In each radial interval, we plot both the median (the
  solid blue line) and the standard deviation of the difference 
  (the grey region), as a function of wavelength.

Overall, all the three curves deviate from the measurements to
some degrees, but in different ways. At $R\le0.5R_e$, the Calzetti curve provides
a best fit to the measurements, as shown in the top-left panel,
while both CCM and SMC curves significantly deviate from the measurements.
At $R>0.5R_e$, the CCM and SMC curves work better than the Calzetti
one, and the SMC appears to behave slightly better than the CCM.
As pointed out above, the selective attenuation curves have fixed values
at $\lambda_B$ and $\lambda_V$. Thus, the wavelength-dependence 
shown in the figure may change if color excess in other two bands 
is used to normalize the curves.

In Fig.~\ref{fig:curve_rms} we further examine the deviation of the
three curves relative to our measurements by plotting the histograms
of the relative difference of $E(B-V)$ (upper panels) and the rms
deviation of the curves (lower panels). Consistent with results shown 
in the previous figure, the Calzetti curve shows the 
smallest deviation at $R\le0.5R_e$, with the distribution of 
$\Delta E(B-V)/E(B-V)$ centered at around zero and a rms 
less than 5\%. At $R>R_e$, the CCM and SMC curves are better than 
the Calzetti curve, with smaller $\Delta E(B-V)/E(B-V)$ on average, 
while the three curves show similar rms in deviation.
These results suggest that none of the three curves can
universally describe the measured curves at all radii. The inner
region of galaxies seems to prefer a Calzetti-like curve 
while the outer region prefers a SMC or Milky Way-like curve.

\section{summary and discussion} \label{sec:summary}

We have developed a method to estimate the relative dust attenuation
curve, $A(\lambda)-A_V$ (see
equation~\ref{eq:fit5}), which is the dust attenuation
as a function of wavelength relative to that
at the fixed wavelength $\lambda_V$. For an observed spectrum 
of a galaxy or a specific region within a galaxy, our method 
first decomposes the spectrum into a small-scale component, 
$S$, and a large-scale component, $L$, by adopting a moving box 
average method. Assuming all the stellar populations 
that contribute to the observed spectrum have the same 
attenuation given by $\Att(\lambda)$, we are able to show that the
ratio of the two components, $R(\lambda)={F_S(\lambda)}/{F_L(\lambda)}$ (see equation~\ref{eq:r_att}), 
is free of dust attenuation. The observed $R(\lambda)$ can be modeled
with a given theoretical library of simple  stellar populations (SSPs), 
without the need for modeling the effect of dust attenuation in 
the fitting. The dust-free intrinsic spectrum is then reconstructed 
from the SSPs according to the corresponding coefficients that 
form the best-fit to $R(\lambda)$. Finally, the relative dust attenuation 
curve is obtained by comparing the observed spectrum with the 
`dust free' model spectrum.

We have performed extensive tests of our method on a set of mock
spectra covering wide ranges in stellar age and metallicity, as well
as $E(B-V)$ and spectral signal-to-noise ratios (SNRs). These tests
show that both the $E(B-V)$ and the relative dust attenuation curve
can be well recovered as long as $E(B-V)>0.05$  and $SNR>5$. At lower
$E(B-V)$ and smaller SNRs, on average, our method tends to
overestimate $E(B-V)$ by $\lesssim 0.03$  magnitude (see
Fig.~\ref{fig:mock}).
  We have also tested our method on more realistic 
  mock spectra using continuous star formation histories and
  including emission lines, finding no significant 
  change in our results
  (see Appendix~\ref{app:mock} and Fig.~\ref{fig:sfh_emls}).

We have applied our  method to the integral
field spectroscopy (IFS) of 15 edge-on galaxies from the ongoing MaNGA
survey, obtaining both the  two-dimensional map and radial profile of
$E(B-V)$ for each galaxy,  as well as the spatially-resolved relative
dust attenuation curve. Using the $E(B-V)$ maps we have corrected the
effect of dust attenuation on the SDSS $gri$ image of these galaxies.
The dust lanes in the original images become 
invisible in the images reconstructed from the attenuation-corrected 
spectra, and the vertical brightness profiles in a given band become
almost symmetric and can be well described by a simple model proposed 
for the disk vertical structure. 

We have compared the estimated dust attenuation
  curves of the 15 MaNGA galaxies with three commonly-used
  empirical models: the Calzetti attenuation curve for local starburst
  galaxies from \citet{2000ApJ...533..682C}, the CCM extinction curve
  for Milky Way from \citet{1989ApJ...345..245C}, and the SMC
  extinction curve for the Small Magellanic Cloud from
  \citet{2003ApJ...594..279G}. 
Although small in number, the MaNGA galaxies in our sample 
already show variations in the slope of dust attenuation curves 
(see the rightmost panels in
Fig.~\ref{fig:manga}), implying that the attenuation curves 
are not universally described by a single curve.
Figs.~\ref{fig:curve_bins} and \ref{fig:curve_rms}
show that the Calzetti curve describe well the attenuation
in the inner region ($R\le0.5R_e$) of the MaNGA galaxies, while
a SMC or Milky Way-like curve works better for the outer region.
This result may be attributed to the negative gradient of star formation
rate observed in spiral galaxies. In this case,
the inner region has stronger star formation than the other region,
and is better described by a Calzetti curve which was initially
proposed for local starburst galaxies.

Once the radial dependence is taken into account, i.e. by adopting 
a Calzeti curve for inner regions and a SMC or Milky Way curve for 
outer regions, the deviations in $E(B-V)$ of the model curves from 
the measured ones are relatively small, with a median of $\sim 10\%$ 
and a rms less than $\sim10\%$ (see Fig.~\ref{fig:curve_rms}).
This result is probably not
unexpected given both the similarity of the different
attenuation models and the relatively weak wavelength-dependence
of the attenuation in the optical band.  It is
known that the various attenuation models differ more significantly in
the ultraviolet. In particular, the CCM model presents a bump at
around 2175\AA, which is absent in the  Calzetti and SMC curves. Therefore,
previous observational studies of dust attenuation curves have mainly
relied on broad-band spectral energy distribution (SED) including UV
bands \citep[e.g.,][]{2016ApJ...827...20S, 2017ApJ...840..109B,
  2017ApJ...851...90B,  2018ApJ...859...11S, 2018MNRAS.475.2363T,
  2019MNRAS.486..743D, 2019ApJ...872...23S}. In many cases, these
studies have also found evidence in support of variations in the slope
of attenuation curves and/or the strength of the UV bump.  In a recent
theoretical paper \citep{2018ApJ...869...70N}, a model for the origin
of such variations was developed, in which the variation in the
attenuation curve slope depends primarily on complexities in the
star-dust geometry, while the bump strength is primarily influenced by
the fraction of unobscured O and B stars. In principle,  if applied to
galactic spectra covering the UV band, our method should be able to
provide an independent way of quantifying the variations in the
attenuation curve and the UV bump. Next-generation large spectroscopic
surveys such as the Prime Focus Spectrograph
surveys \citep[PFS;][]{2014PASJ...66R...1T} will obtain
high-quality spectra for hundreds of thousands of galaxies
at $1\lesssim z\lesssim2$, for which the rest-frame UV bump
will be well covered in the observed spectrum. We expect to apply
our method to those galaxies, thus directly deriving the selective
dust attenuation curves for high-$z$ galaxies.

  One important advantage of our method is that the estimated dust
attenuation curve is independent of the shape of theoretical dust
attenuation curves. In other words, this method provides a way to
break the known degeneracy of dust attenuation with other properties
of the stellar population in a galaxy, particularly stellar age
and metallicity. Therefore, we can expect to have better
constraints on the stellar age and metallicity by performing 
full spectral fitting to the {\em `dust-free'} spectrum obtained using
our method. In Appendix~\ref{app:degeneracy} we present a test of this
hypothesis on a set of mock spectra, using a Bayesian approach to
explore the potential degeneracy of model parameters. As can
be seen there, the uncertainties in both the age and metallicity are 
indeed reduced significantly by using the estimated dust attenuation 
curve before the spectral fitting. We will come back to this point
and further explore the potential of our method in improving 
stellar population synthesis techniques.

As emphasized, one limitation of our method is that the estimated dust
attenuation curves are relative, but not absolute. 
Equation~\ref{eq:rv} indicates that, if one were to obtain the
total (absolute) dust attenuation at given $\lambda$, it is necessary
to know the value of $R_V$ (or, more generally, the total
attenuation at any given wavelength that falls in the wavelength
range covered).  For instance, the $R_V$ of the Calzetti dust curve
is 4.05, with a dispersion of 0.8  \citep{2000ApJ...533..682C}.
In reality, however, $R_V$ may vary from galaxy to
  galaxy, and from region to region within a galaxy.  Different
  galaxies or regions may have the same shape of attenuation curve,
  but different absolute attenuation values at a given wavelength due to
  the variation of $R_V$
  \citep[e.g.,][]{2015ApJ...806..259R,2018ApJ...859...11S}. One
  plausible way to determine $R_V$, thus the absolute
  attenuation curve, is to include data in near-infrared (NIR),
  e.g. photometry in $J$, $H$ and $K_s$ bands from existing NIR 
  imaging surveys, where dust attenuation becomes negligible. For a given
  $R_V$, in practice, one may predict the apparent magnitudes in the NIR 
  from the dust-free spectrum, $F^{\mbox{fit}}(\lambda)$, as obtained from
  fitting the $S/L$ ratio of the observed spectrum  (see
  Eqn.~\ref{eq:fit2}). The value of $R_V$ and the absolute
  attenuation curve can then be derived by comparing the predicted NIR
  magnitudes with the observed ones. We will test and apply this
  idea in the future.

\acknowledgments
    {We're grateful to the anonymous referee whose comments have
      helped improve this paper. 
      This work is supported by the National Key R\&D Program of China
(grant No. 2018YFA0404502, 2018YFA0404503), and the National Science Foundation of
China (grant Nos. 11821303, 11973030, 11673015, 11733004,
11761131004, 11761141012).

Funding for the Sloan Digital Sky Survey IV has been provided by the
Alfred P. Sloan Foundation, the U.S. Department of Energy Office of
Science, and the Participating Institutions. SDSS-IV acknowledges
support and resources from the Center for High-Performance Computing
at the University of Utah. The SDSS web site is www.sdss.org.

SDSS-IV is managed by the Astrophysical Research Consortium for the
Participating Institutions of the SDSS Collaboration including the
Brazilian Participation Group, the Carnegie Institution for Science,
Carnegie Mellon University, the Chilean Participation Group, the
French Participation Group, Harvard-Smithsonian Center for
Astrophysics, Instituto de Astrof\'isica de Canarias, The Johns
Hopkins University, Kavli Institute for the Physics and Mathematics of
the Universe (IPMU) / University of Tokyo, Lawrence Berkeley National
Laboratory, Leibniz Institut f\"ur Astrophysik Potsdam (AIP),
Max-Planck-Institut f\"ur Astronomie (MPIA Heidelberg),
Max-Planck-Institut f\"ur Astrophysik (MPA Garching),
Max-Planck-Institut f\"ur Extraterrestrische Physik (MPE), National
Astronomical Observatories of China, New Mexico State University,New
York University, University of Notre Dame, Observat\'ario Nacional /
MCTI, The Ohio State University, Pennsylvania State University,
Shanghai Astronomical Observatory, United Kingdom Participation Group,
Universidad Nacional Aut\'onoma de M\'exico, University of Arizona,
University of Colorado Boulder, University of Oxford, University of
Portsmouth, University of Utah, University of Virginia, University of
Washington, University of Wisconsin, Vanderbilt University, and Yale
University.}

\appendix

\begin{figure*}
  \centering
    \includegraphics[width=1\textwidth]{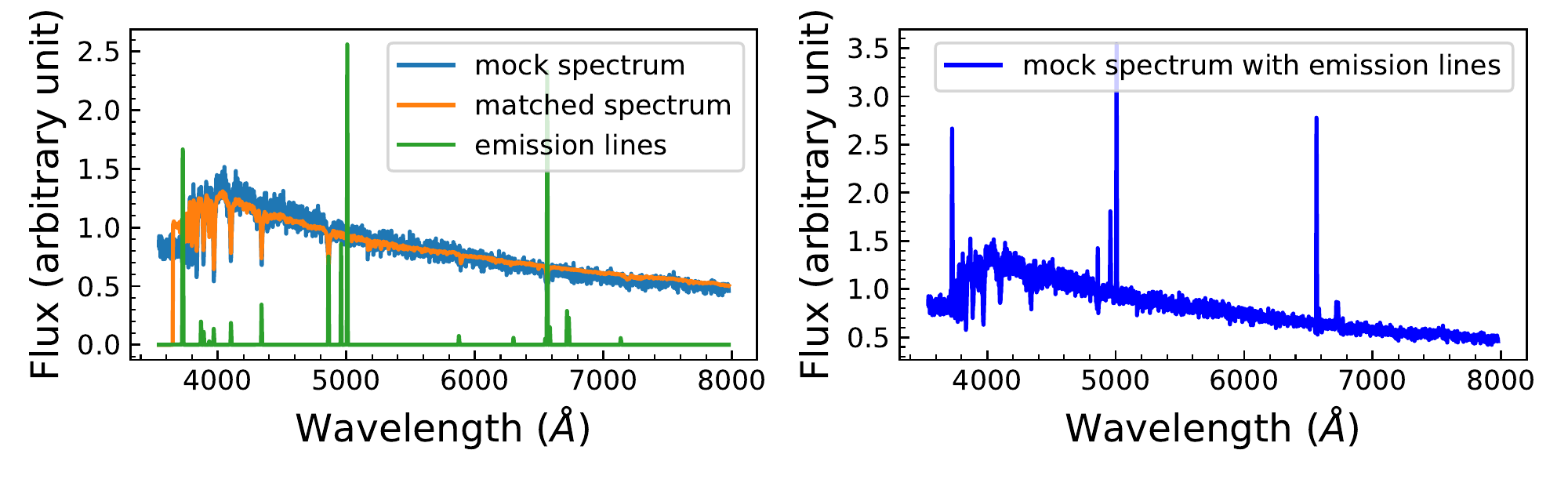}
    \caption{An example of mock spectrum with emission lines. In the left panel,
      light blue line is the mock spectrum without emission lines.
      Orange line is the best-fit stellar spectrum of the {\em matched}
      spectrum in the MaNGA sample (SDSS/DR14). Green line shows the
      observed emission lines in the matched MaNGA spectrum.
      The right-hand panel shows the final mock spectrum with
      emission lines, which is actually the sum of the blue and green lines
      in the left panel.}
    \label{fig:mock_emlines}
\end{figure*}

\begin{figure*}
  \centering
    \includegraphics[width=1\textwidth]{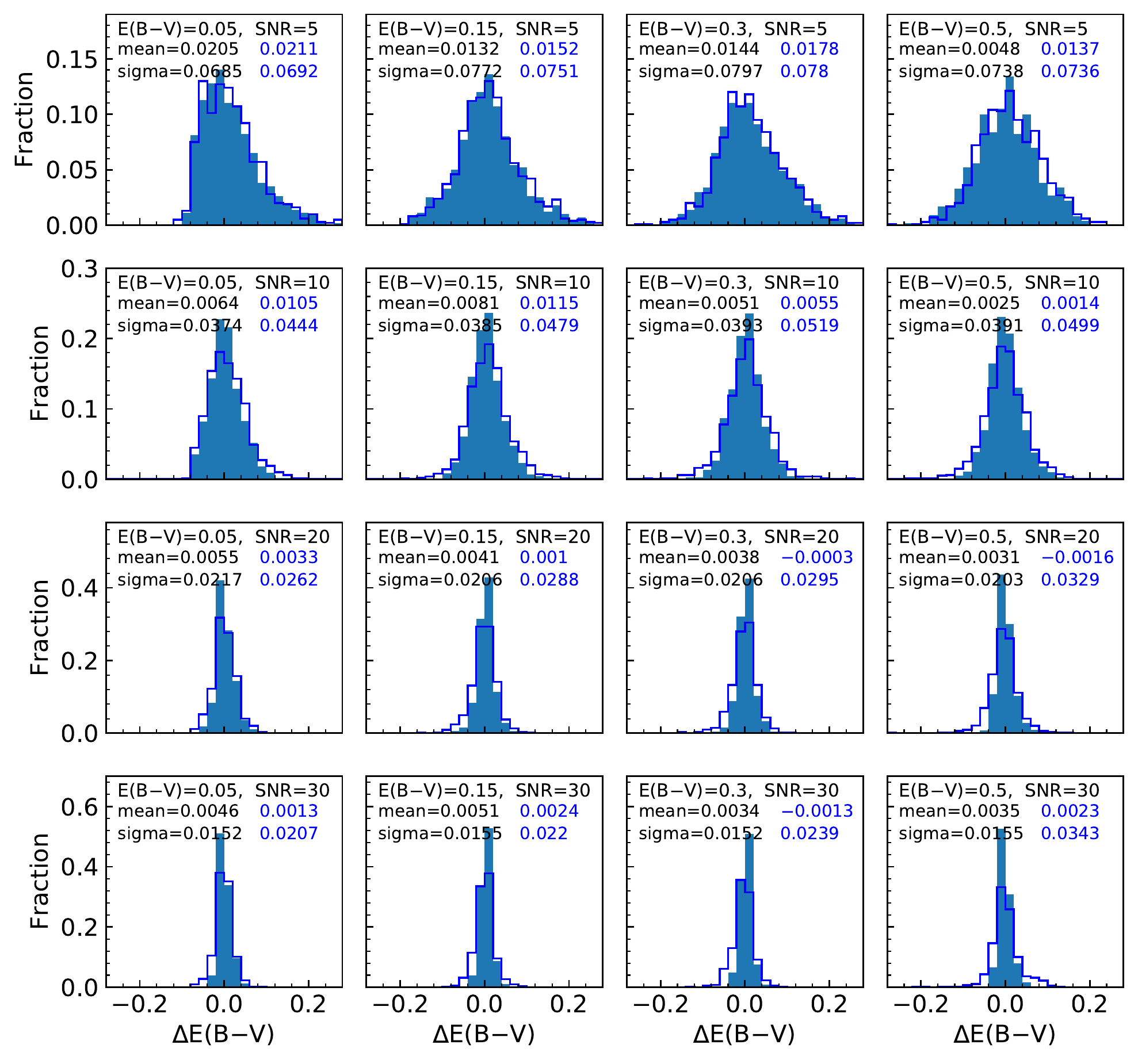}
    \caption{Tests on mock spectra generated with continuous SFH,
      with (unfilled histograms) or without (filled histograms) including
      emission lines. Panels from left to right correspond to different
      $E(B-V)$, and panels from top to bottom correspond to different
      signal-to-noise ratios (SNRs), as indicated. In each panel,
      the histograms show the distribution of the difference between
      the output and input value of $E(B-V)$, referred to as $\Delta E(B-V)$.
      Indicated in each panel are the mean value and standard deviation 
      of $\Delta E(B-V)$ for both the filled (black) and unfilled (blue) histograms.
      }
    \label{fig:sfh_emls}
\end{figure*}

  \section{Tests with more realistic mock spectra}
  \label{app:mock}

  In \S\ref{sec:mock} we have tested our method of estimating dust
  attenuation curves on a set of mock spectra generated by
  linearly combining discrete simple stellar populations (SSPs)
  covering a wide range of stellar age and metallicity. In a real
  galaxy, however, stars are expected to form over an extended period 
  of time with a continuous star formation history (SFH). In
  addition, observed spectra of galaxies often present emission lines
  produced by ionized gas.  Here we present additional
  analyses by testing our method on a set of more realistic mock
  spectra with both continuous SFHs and emission lines. As we will
  show below, even with these additional complications,  
  our method is still capable of recovering the input dust 
  attenuation curves reliably and accurately.
  
\subsection{Generating the mock spectra}

For the SFH, we adopt the widely-used $\tau$-model in which the star
formation rate declines with time exponentially, 
\begin{equation}
  \Psi(t)=\frac{1}{\tau}\frac{e^{-(t-t_i)/\tau}}{1-e^{-(t_0-t_i)/\tau}},
\end{equation}
where $\tau$ is the $e$-folding timescale, $t_i$ and $t_0$ are the
starting time and the present time, respectively, so that
$(t_0-t_i)$ is the total duration of the SFH. A small $\tau$ and a
large $(t_0-t_i)$ thus correspond to a short duration of 
star formation at early times, which may be suitable for 
a galaxy dominated by old stellar populations. In contrast, a 
long $e$-folding timescale and a small
value of $(t_0-t_i)$ are suitable for young galaxies with steadily
declining star formation. For
simplicity, we adopt $t_0=14$ Gyr, and generate 1000 SFHs by randomly
selecting $\tau$ and $t_i$ over the ranges of $0.01<\tau<100$Gyr and
$0<t_i<13.9$Gyr, respectively. For a given SFH, we first convolve it 
with simple stellar populations (SSPs) given by the BC03 library
(\citealt{2003MNRAS.344.1000B}), which are obtained using the Padova
1994 evolutionary tracks and a Chabrier initial mass function
\citep{2003PASP..115..763C}. This gives a noise-free composite
spectrum, which is then reddened with a Calzetti attenuation curve,
assuming four different color excesses: $E(B-V)$=0.05, 0.15, 0.3 and
0.5. Different levels of Gaussian noise are added to the spectrum
to mimic signal-to-noise ratios of SNR=5, 10, 20 and
30, respectively. In total we have a set of 16,000 spectra that 
cover wide ranges of age, metallicity, SNR and color excess.

Finally, we add emission lines to each of the mock spectra. To do
this, we take the real spectra of MaNGA galaxies in the SDSS/DR14, and
perform full spectral fitting to each spectrum using the public
pipeline \texttt{pPXF}
\citep{2004PASP..116..138C,2017MNRAS.466..798C}.  For each of our mock
spectra, we find a counterpart spectrum from the real spectra of
MaNGA, requiring that their SNRs are similar ($\Delta SNR<5$) and 
that the fitted spectrum and the mock spectrum are matched with each other 
most closely according to $\chi^2$ minimum. The starlight-subtracted
emission lines of the counterpart spectrum are then added to the mock
spectrum.  Fig~\ref{fig:mock_emlines} displays an example of the mock
spectra with emission lines. We have visually examined a considerable
fraction of the mock spectra.  In general, mock spectra with younger
ages present stronger emission lines, while spectra of older ages
present weaker emission lines, consistent with expectations.

\subsection{Test results}

We test our method on the mock spectra both before and after the
emission lines are added. For the mock spectra with emission lines, we
have carefully masked out all emission lines,  following the procedure
described in detail in \citet{2005AJ....129..669L}, before we apply
our method to measure the dust attenuation curve. In short, emission
lines are identified from starlight-subtracted emission-line spectrum,
and line widths are measured by fitting each line with a Gaussian
profile. This procedure is iterated two or three times,
and the line widths so obtained are used to determine the mask window
of each line. 

Fig~\ref{fig:sfh_emls} shows the result of the test. Panels from left
to right correspond to the four different $E(B-V)$ values, and panels from top to
bottom correspond to the four different SNRs. 
In each
panel, the filled histogram shows the distribution of $\Delta E(B-V)$,
the difference between the output and input values of $E(B-V)$, for the
1000 mock spectra with no emission lines, while the unfilled histogram
shows the result for the same set of 1000 mock spectra but with
emission lines added. The mean and standard deviation of the filled
distribution, as well as the $E(B-V)$ and SNR of the mock spectra
are indicated in each panel. It is encouraging that the results
here are quite similar to those shown in Fig~\ref{fig:mock}. In all cases
except the top two panels in the leftmost column where $E(B-V)=0.05$
and SNR$\leq 10$, the histograms are close to a Gaussian with a mean
around zero and a small value of standard deviation. In addition, the
results remain almost unchanged when emission lines are included, 
although the standard deviation becomes slightly larger.

We therefore conclude that our method provides reliable and unbiased
measurements of $E(B-V)$ for more realistic spectra with continuous
SFHs and emission lines.

\begin{figure*}
  \centering
    \includegraphics[width=1\textwidth]{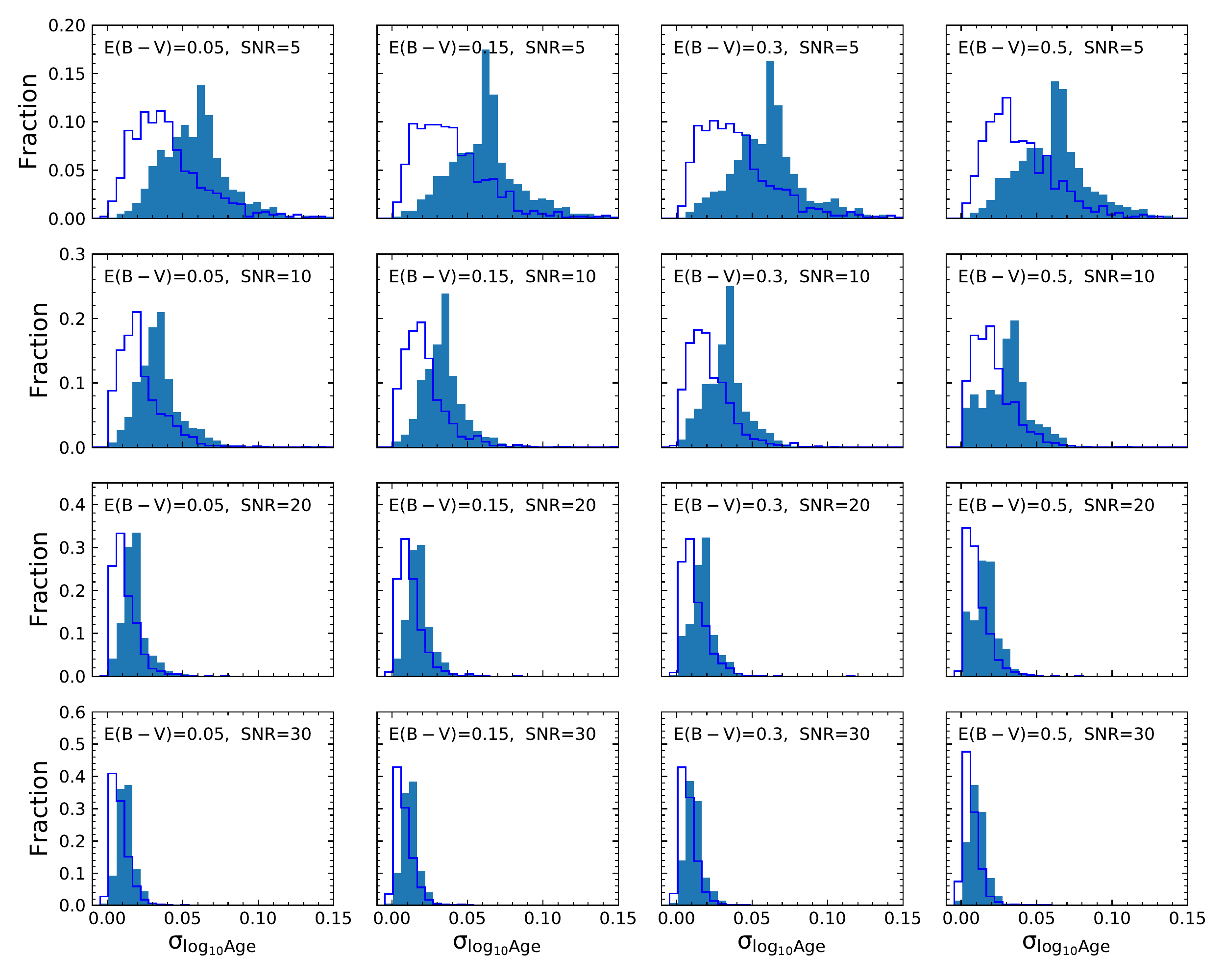}
    \caption{Histograms of uncertainty of stellar age in stellar population
      synthesis (SPS), as quantified by $\sigma_{\log_{10}{\rm Age}}$
      which is the $1-\sigma$ width of the
      posterior distribution of $\log_{10}{\rm Age}$ inferred
      by fitting the mock spectra with our SPS code \texttt{BIGS}.
      Different panels are for different $E(B-V)$ and signal-to-noise
      ratios (SNSs), as indicated. In each panel the filled histogram
      is for the case where $E(B-V)$ is a free parameter and constrained
      simultenously with other stellar population properties during
      the spectral fitting, while the unfilled histogram is for
      the case where the spectrum being considered is corrected for
      dust attenuation using our method before the fitting.
    }
    \label{fig:sigma_age}
\end{figure*}

\begin{figure*}
  \centering
    \includegraphics[width=1\textwidth]{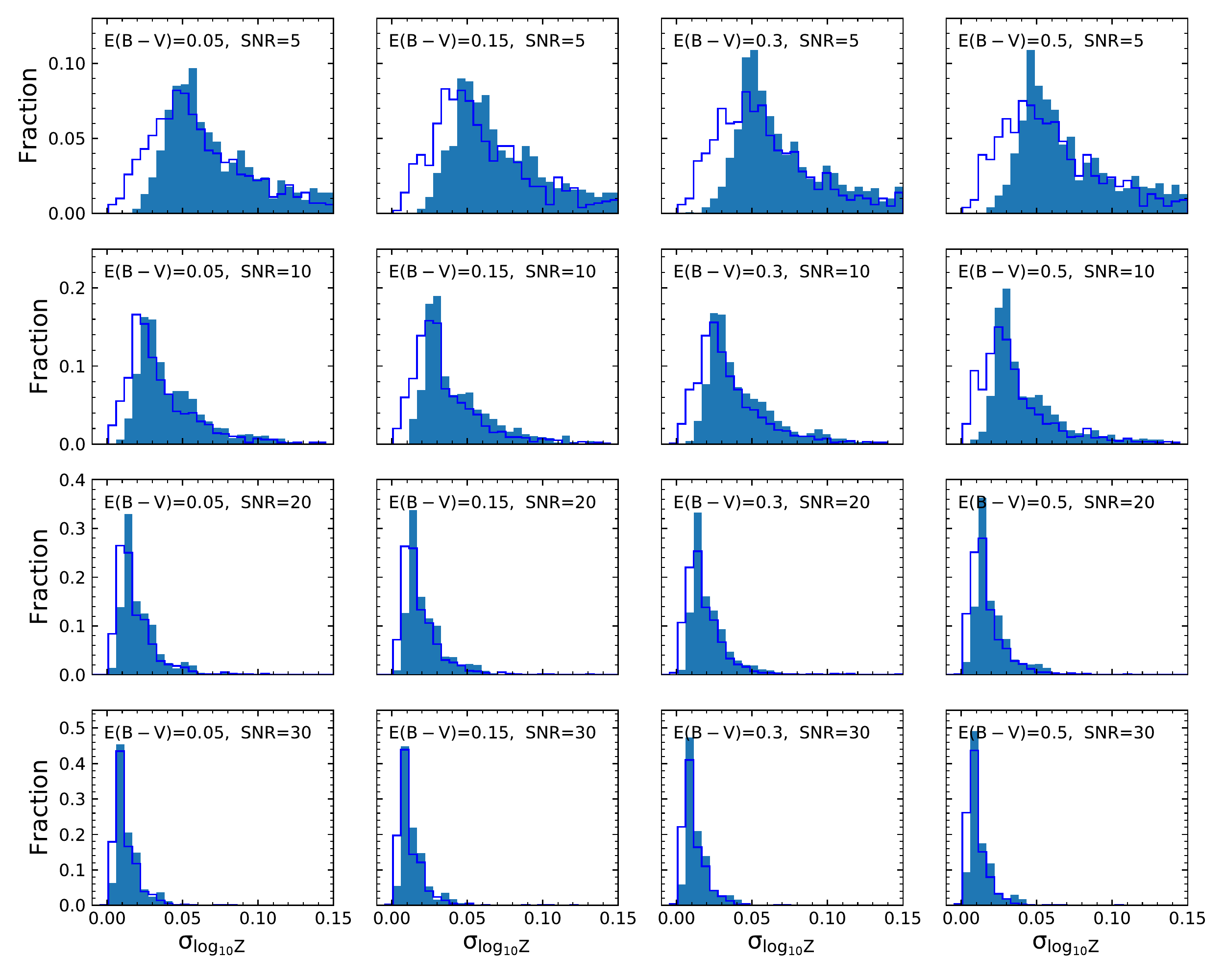}
    \caption{Same as Fig.~\ref{fig:sigma_age} but for stellar
      metallicity instead of stellar age.}
    \label{fig:sigma_met}
\end{figure*}

  \section{Alleviating the degeneracy between dust and stellar populations}
  \label{app:degeneracy}

As emphasized in the main text, one important advantage of our method
is that the estimated (relative) dust attenuation curves are
independent of the shape of theoretical attenuation curves. The
method provides a reliable and direct way to estimate the attenuation
curve and to correct the effect of attenuation for the
observed spectrum, thereby avoiding dust attenuation 
in stellar population synthesis (SPS) modeling. 
In principle, this should also help break the degeneracy 
between dust, age and metallicity in observed spectra. 
Here we present a test of this hypothesis using the same set of mock 
spectra generated in the previous section. For this purpose, we use our 
recently developed SPS code, Bayesian Inference of Galactic
Spectra \citep[BIGS,][]{2019MNRAS.485.5256Z}. The code performs full
spectral fitting to the observed spectrum of a galaxy (or a region
within it), and derives Bayesian inferences of age and metallicty
along with other properties of its stellar population. In particular, 
the Bayesian approach provides a statistically rigorous way to explore
potential degeneracy among model parameters.

For simplicity, we consider only mock spectra with emission lines.
We first apply \texttt{BIGS} to each of the mock spectra and obtain 
the corresponding posterior distribution of stellar population 
properties, such as age, metallicity and $E(B-V)$. Next, we apply 
our method to estimate the dust attenuation and to correct the 
effect of attenuation. Finally, we use
\texttt{BIGS} to fit the {\em `dust-free'} spectrum. 
Figs.~\ref{fig:sigma_age} and \ref{fig:sigma_met} show the
$1-\sigma$ width of the posterior distribution of stellar age
($\sigma_{\log_{10}{\rm Age}}$) and metallicity ($\sigma_{\log_{10}{\rm Z}}$),
for different $E(B-V)$ and SNR. In each panel, the filled
histogram is the result obtained from applying \texttt{BIGS} directly 
to the mock spectra, with dust attenuation treated as a free parameter
during the fitting. The unfilled histogram is the result of fitting
the spectra in which dust attenuation is estimated and
corrected using our method. As one can see, in all cases 
the inference uncertainties for both of the stellar parameters are reduced 
when the dust attenuation is corrected before the fitting. 
The improvement is more pronounced for lower SNRs, but with week 
or no dependence on $E(B-V)$. These results confirm the above hypothesis 
and motivate applications of our method of modeling attenuation  
and \texttt{BIGS} to real spectra of galaxies
(N. Li et al., in prep.).

\bibliography{main}{}
\bibliographystyle{aasjournal}



\end{document}